# 3D Lagrangian turbulent diffusion of dust grains in a protoplanetary disk: method and first applications

Sébastien CHARNOZ[1*]

Laure FOUCHET[2]

Jérôme ALEON [3]

Manuel MOREIRA[4]

(1) Laboratoire AIM, Université Paris Diderot /CEA/CNRS  UMR 7158
    91191 Gif sur Yvette cedex FRANCE

(2) Physikalisches Institut, Universität Bern, CH-3012 Bern, SWITZERLAND

(3) Centre de Spectrométrie Nucléaire et de Spectrométrie de Masse, CNRS/IN2P3, Université Paris Sud 11, Bâtiment 104, 91405 Orsay Campus, FRANCE

(4) Institut de Physique du Globe de Paris, Université Paris-Diderot, UMR CNRS 7154, 1 rue Jussieu, 75238 Paris cedex 05, FRANCE

(*) To whom correspondence should be addressed: charnoz@cea.fr

**ABSTRACT***:*

In order to understand how the chemical and isotopic compositions of dust grains in a gaseous turbulent protoplanetary disk are altered during their journey in the disk, it is important to determine their individual trajectories. We study here the dust-diffusive transport using lagrangian numerical simulations using the the popular "turbulent diffusion" formalism. However it is naturally expressed in an Eulerian form, which does not allow the trajectories of individual particles to be studied. We present a simple stochastic and physically justified procedure for modeling turbulent diffusion in a Lagrangian form that overcomes these difficulties. We show that a net diffusive flux F of the dust appears and that it is proportional to the gas density (ρ) gradient and the dust diffusion coefficient $D_d$: (F=$D_d$/ρ×grad(ρ)). It induces an inward transport of dust in the disk's midplane, while favoring outward transport in the disk's upper layers. We present tests and applications comparing dust diffusion in the midplane and upper layers as well as sample trajectories of particles with different sizes. We also discuss potential applications for cosmochemistry and SPH codes.

# INTRODUCTION

The transport of solids in turbulent, gaseous protoplanetary disks involves highly diverse physical processes such as gas drag, turbulence, photophoretic gas pressure and radiation pressure. Modeling these processes is important and necessary as an increasing amount of observational data testifies that, both in our own Solar System and in protoplanetary disks orbiting distant stars, large-scale transport of solids occurs over tenths of astronomical units (AU). Samples from the comet 81P/Wild2 (Brownlee et al. 2006; Zolensky et al. 2006; Westphal et al. 2009) and observations of comet 9P/Tempel 1 (Lisse et al. 2006) have revealed the presence of crystalline silicates, which may have originated within 2 AU of the Sun and then been transported outwards by some mechanism. In this respect, perhaps the most spectacular result of the *Stardust* mission to comet Wild2 is the discovery of refractory inclusions formed at high temperatures ($\geq 1500$ K) in the innermost regions of the solar protoplanetary disk (<< 1 AU) in a comet originating in the Kuiper Belt (e.g., Zolensky et al. 2006; Simon *et al.* 2008; Matzel et al. 2010). Similar observations made with the Spitzer space-telescope have revealed that crystalline silicates are also found in T-Tauri disks with crystalline-to-amorphous silicate ratios varying greatly from one disk to another (Van Boekel et al. 2005; Olofson et al.; 2009, Watson et al. 2005). These results suggest that global dust transport processes are indeed active, but that their nature and efficiency may differ significantly from one disk to another. Since dust settling is expected to modify the photometric appearance of a disk (Dullemond & Dominik 2004), observations can reveal the disk structure and signs of transport. For example, in the GG Tauri circumbinary disk (Duchêne et al. 2004; Pinte et al. 2007), multi-wavelength observations have revealed a radial size-sorting of dust particles at least qualitatively consistent with the radial migration induced by gas drag.

Many studies have addressed different physical aspects of dust transport (for radiation pressure, see e.g., Vinkovic 2009; photophoresis, see e.g., Krauss & Wurm 2005; stellar wind, see e.g., Shu et al. 2001; turbulent diffusion, see e.g., Gail 2001 and Ciesla 2009; photoevaporation, see e.g., Alexander & Armitage 2007). These studies use various tools and formalisms, either Eulerian or Lagrangian, which are most often designed to study one specific aspect of transport. For example, multi-fluid simulations are appropriate for describing the turbulent transport of fine dust (see e.g., Fromang & Nelson 2009) but very computationally demanding. At the opposite end, populations of large particles (which are less collisional) are not accurately described by a fluid approach and Lagrangian approaches seem to be more adapted given that they track individual trajectories (see e.g., Johansen & Youdin 2007; Johansen et al. 2007).

It would thus be useful to build a 3D modular code that allows for a simple coupling of the different physical processes of dust transport. As a first step, the present paper describes a 3D Lagrangian dust transport code in which the equation of motion is numerically solved with as few approximations as possible. While the motion of small dust particles that are tightly coupled to gas is analytically well-known, the motion of particles loosely coupled to gas requires direct integration. Here we focus on the motion of dust particles under gas drag and turbulent diffusion. Other processes will be incorporated into future work. Because the effect of gas drag in a laminar disk using a Lagrangian description is largely documented (see e.g., Barrière-Fouchet et al. 2005), this paper focuses mainly on the most difficult aspect: the inclusion of turbulent diffusion in a Lagrangian code. We describe below the various features of this code.

*A Lagrangian approach*

A Lagrangian code treats each particle individually, which means that it is not well suited for representing a fluid system where the mean free path is short. It is, however, well designed for tracking point-like particles in a gas disk that have very few pairwise interactions. One very attractive advantage of this approach is that it affords the possibility of tracking individual particle trajectories, and thus, of reconstructing the thermodynamical history of particles in a protoplanetary disk environment. This should have many applications for cosmochemistry. For instance, chondrules and refractory inclusions, which are the main millimeter- to centimeter-sized components of primitive chondritic meteorites, are thought to have undergone a complex multi-stage evolution during the first 2-3 Myr of Solar System history, before being incorporated into asteroidal or cometary planetesimals at various heliocentric distances. Chondrules (Mg-Fe-rich) and refractory inclusions (also known as Calcium-Aluminium Inclusions, CAIs hereafter) were formed in the inner solar system and experienced multiple heating and irradiation events before, or during, their transport to the region they were finally incorporated into a meteorite. As they are the major building blocks of rocky planetesimals, tracking their thermodynamical history is a key information apjto deciphering the physics and chemistry of planetary formation. Similarly, the history of frozen volatile-rich dust grains from outer solar system regions is essential to understanding the solar protoplanetary disk and to unraveling the origin of planetary volatiles, such as water or organic molecules with prebiotic potential.

*Turbulent diffusion*

The disk is expected to be MRI turbulent, inducing an efficient mixing of the dust component and the gas component. A direct approach would be to couple a particle-based dust transport code with a 3D MHD simulation of a turbulent gas disk (as in Fromang & Nelson 2005; Johansen & Youdin 2007; Johansen et al. 2007). However, while conceptually simple, MHD simulations remain very computationally demanding, and consequently are currently limited to a few thousands orbits at most. For this reason, we turn our attention to a simplified turbulence model, the popular *turbulent diffusion model,* which mimics turbulent transport as a diffusive process through a Brownian motion with an efficiency parameter: the dust diffusion coefficient $D_d$. This diffusion coefficient is thought to be comparable in magnitude to the turbulent viscosity coefficient. More recently, Fromang and Nelson (2009) have shown that $D_d$ may increase with the distance from the midplane. The description of turbulence as a diffusive process, though not fully accurate, is widely used and underpinned by a vast literature. We have therefore built on earlier studies to develop an efficient Lagrangian code. For example, Lagrangian diffusion has been extensively used in environmental studies for the transport of air- and ocean-borne pollutants (see Wilson & Sawford 1996 for a review). In the planetary science literature, several models couple gas drag and turbulent diffusion within a Lagrangian framework, but this is either in a form not adapted to large particles (which decouple from the gas and undergo oscillations) or limited to some specific prescription of the gas density field. For example, in Ciesla (2010) and in Hughes and Armitage (2010), a 1D stochastic diffusion model (partly similar to ours; see Section 2) is applied, but the treatment of gas density variations is different. Our method is readily applicable to any 3D gas disk and thus much more general than the one used in Ciesla (2010). It is also more physically justified than in Hughes and Armitage's (2010) model, which is mainly empirical. Another major difference is our use of an implicit solver to integrate the dust motion, which allows us to accurately follow any range of particle size, from the most tightly gas-coupled particles, such as polycyclic aromatic hydrocarbons (PAH), to those totally decoupled from the gas, such as kilometer-sized planetesimals (see the examples in Section 4.2). We will see, in particular, that proper implementation of turbulent diffusion is not straightforward, with one frequently encountered problem being that of satisfying the "good mixing condition", i.e. reaching an asymptotic state in which dust is well mixed with the gas for any gas spatial density distribution, in accordance with the second law of Thermodynamics. This requires a special treatment of diffusion in a Lagrangian approach and constitutes the core of this paper. The case of a non-constant diffusion is also addressed.

*Three-dimensional system*

Another useful physical aspect of our code is that we consider dust motion in three dimensions. Transport of dust to altitudes high above the midplane is expected due to the efficient vertical diffusion induced by turbulence. Most studies treat radial mixing in the protoplanetary disk through a 1D approach (see e.g., Gail 2001; Brauer et al. 2008; Hughes & Armitage 2010). Yet, 2D models that explicitly treat the vertical motion of dust particles (see e.g., Takeuchi & Lin 2002; Dullemond & Dominik 2004; Ciesla 2007; Tscharnuter & Gail 2007) show that the disk is stratified, which may have an impact on global transport. For example, Takeuchi and Lin (2002) show that the radial velocity of dust depends quadratically on Z inducing an *outward* gas drag in the disk's upper layers (for above ~1.5 pressure scale heights, see Takeuchi & Lin 2002, Eq. [11] and Eq. [17]). This depends on the gas density profile only, and is thus a robust result. We shall also see that radial diffusion is more effective at altitudes far from the midplane, due to the shallower slope of the radial density gradient for Z>0 (see Section 3.2). The importance of including the vertical dimension is also emphasized by Bai and Stone (2010) in the context of dust transport in a dead zone. Thus, in order to ensure that the approach remains as general as possible, it is important to incorporate the vertical dimension into the system and integrate the motion of each particle with the fewest approximations possible. In the present paper, the azimuthal direction is also explicitly included. However, as there is no planet for the moment, the disk remains azimuthally symmetric, while the system is intrinsically evolved in 3D.

For practical use, this code will be referred to as LIDT3D (for Lagrangian Implicit Dust Transport in 3D).

The paper is organized as follows: in Section 2 we describe the procedure used to introduce turbulent diffusion into a Lagrangian form. It should be noted that the gas disk considered in this paper is a simple and non-evolving parameterized gaseous disk (as in Takeuchi & Lin 2002) in order to test the dust transport algorithm. This algorithm, however, is independent of the choice of disk and can be readily extended to any disk sampled on a 3D grid. In Section 3, we present several tests aimed at reproducing known results about turbulent diffusion in a gaseous disk and in Section 4 we discuss some applications to compare dust diffusion in two dimensions (at the disk midplane only) and in three dimension and to show individual trajectories of different-sized dust grains extracted from the simulations.

## 2. Numerical implementation of turbulent diffusion

### 2.1 Basic concepts and definitions

In a Lagrangian code, the motion of an individual dust particle is described using the classical Newtonian formalism:

$$\frac{d\vec{v}}{dt} = \frac{\vec{F}_*}{m} - \frac{\vec{v} - \vec{v}_g}{\tau} , \qquad (1)$$

where $\vec{F}_*$ is the gravitational force of the central star, the second term is the gas drag force, $v$ is the particle's velocity, $v_g$ is the gas velocity and m is the particle's mass. The dust stopping time $\tau$ is in the Epstein regime:

$$\tau = \frac{a \rho_s}{\rho C_s} , \qquad (2)$$

where *a* is the dust grain radius, $\rho_s$ is the dust material density, ρ is the gas density and $C_s$ is the local sound velocity. Particles with sizes larger than the mean-free-path of gas molecules may be in the "Stokes Regime", where the exact expression for τ depends on the gas Reynolds number (see Eq.[10] of Birnstiel et al. 2010). However, our discussion and the method described here do not closely depend on the expression for the stopping time τ. We introduce the Stokes number St, which is the particle coupling time τ divided by the eddy turnover time $\tau_{ed}$ (St = $\tau/\tau_{ed}$). $\tau_{ed}$ is about $1/\Omega_k$ (see e.g., Fromang & Papaloizou 2006), with $\Omega_k$ representing the local Keplerian frequency ($\Omega_k=(GM_*/r^3)^{1/2}$), where $M_*$,G and r are the star's mass, the universal gravitational constant, and the distance to the star projected along the disk midplane, respectively, such that St~$\tau\Omega_k$. In the laminar case, once the gas velocity field is known by applying, for example, a numerical method like used by Tscharnuter and Gail (2007), or an analytical model like Takeuchi & Lin 2002 (as is used here), Eq. (1) is easy to solve numerically, in the absence of turbulence. However, since gas is turbulent, the gas velocity field is highly complex, then, it is more convenient to introduce a turbulent diffusion model into this equation to take the stochastic motion of the gas into account (see e.g., Hinze 1959; Tchen 1947; Youdin & Lithwick 2007). We first transform Eq. (1) by decomposing $v_g$ into a mean-field contribution <$v_g$> plus a turbulent fluctuation **δ$v_g$**, such that **$v_g$**=< $v_g$ >+ **δ$v_g$**, so that Eq. (1) is re-written:

$$\frac{d\vec{v}}{dt} = \frac{\vec{F}_*}{m} - \frac{\vec{v}-<\vec{v}_g>}{\tau} + \frac{\vec{\delta v_g}}{\tau}, \qquad (3)$$

where **δ$v_g$**/τ is the acceleration of dust induced by the turbulent gas velocity field that induces a random walk (or turbulent diffusion) of the particle, consistent with the "turbulent diffusion" model. We assume that <**$v_g$**> can be identified with the unperturbed laminar flow. This gas flow can either be numerically computed as in Ciesla (2009) or Tscharnuter and Gail (2007) or taken from analytical models as in Takeuchi and Lin (2002). Whereas the former method is more versatile, we choose to adopt the latter in the present paper for the sake of simplicity and also to enable us to test our dust transport model against analytical results. Yet, the method for including turbulent diffusion that we present below is not dependent on this choice and a gas disk sampled on a 2D or 3D grid can be easily implemented to provide the <**$v_g$**> field.

We introduce **δ$v_T$** and **δ$r_T$** (kicks in velocity and position) defined as **δ$r_T$**=dt×**δ$v_{Tv}$**, where dt is the time-step. They are decomposed into the following elements by integrating Eq. (3):

$$\vec{\delta v} = \vec{\delta v_M} + \vec{\delta v_T} \qquad (4)$$

$$\vec{\delta r} = \vec{\delta r_M} + \vec{\delta r_T}, \qquad (5)$$

where **δ$v_M$** and **δ$v_T$** are velocity increments due to the central star gravity and gas drag with the disk mean velocity field (**δ$v_M$**) and gas drag with the turbulent velocity field (**δ$v_T$**). As regards positions, **δ$r_M$** is position increments due to initial velocity plus the star's gravity field and gas drag with the mean flow. **δ$r_T$** corresponds to the position increment due to turbulent diffusion that induces a particle's random walk. The turbulent random walk of dust is entirely contained in the **δ$r_T$** and **δ$v_T$** terms, as discussed in Section 2.2. The term **δ$r_M$** is obtained by direct numerical integration of the equation:

$$\vec{\delta r_M} = \int_t^{t+dt} v(t)dt \qquad (6)$$

with

$$\frac{d\vec{v}}{dt} = \frac{\vec{F}_*}{m} - \frac{\vec{v} - <\vec{v}_g>}{\tau} \quad . \tag{7}$$

Eq. (6) and Eq. (7) can be solved using a variety of implicit or explicit variable-order methods (see e.g., Press et al. 1992). An implicit method is highly recommended here given that Eq. (1) is a stiff equation due to the two very different timescales at play: the stopping timescale $\tau$ and the Keplerian orbital timescale $T_k$. The smallest particles have a very short value of $\tau$ (meaning that they are tightly coupled to the gas) that may be much smaller than $T_k$: for example, 0.1 micron and 1mm particles at 1 AU in the minimum-mass solar nebula have $\tau/T_k$ about $10^{-7}$ and $10^{-3}$, respectively. If an explicit integrator such as the popular explicit fourth-order Runge-Kutta scheme is used, the integration time-step will be limited to a fraction of the gas-coupling timescale $\tau$. It will then be impossible to include the smallest particles (with a small Stokes number) or to integrate them over long timescales.

In the present work we use a Bulirsch-Stoer scheme with a semi-implicit solver (Bader & Deuflhard 1983) as described in Press et al. (1992, Chapter 16.6). This yields excellent results in terms of accuracy and rapidity, at least down to a Stokes number of about $10^{-8}$ when we use the Bulirsch-Stoer extrapolation method up to the eighth order in the case of a laminar flow. However the taking into account of turbulent diffusion (see next section) is only first order accurate due to operator split. In this case the Bulirsch-Stoer scheme is used with a 2$^{nd}$ order intregrator. Due to the adaptive time-step scheme it yields excellent results in terms of stability whereas the intrinsic accuracy is only first order. However extensive tests (see section 4) have shown that our results match precisely analytical expectations even when turbulence is included.

**2.2 The "good mixing" problem**

A diffusion process is usually described by the Fick's law, which relates the diffusive flux of material $F_T$ to the density gradient:

$$\vec{F}_T = -D \cdot \overrightarrow{grad}(\rho) , \tag{8}$$

where D is the diffusion coefficient and $\rho$ is the local material density. For the specific case of dust in the solar nebula, D will be written hereafter $D_d$ and $\rho$ is written $\rho_d$. To mimic a random walk of the particles and obtain a diffusion flux obeying Eq. (8), a well-known method is to express $\delta r_T$ or $\delta v_T$ as Gaussian random variables with 0 mean and a variance dependant on the diffusion coefficient and the time-step (see e.g., Wilson & Sawford 1995; Hughes & Armitage 2010 and Annex 1 of the present paper). We now go on to consider a 1D variable only, but the procedure is readily generalized to three dimensions. We express the variance of $\delta r_T$ as $\sigma_r^2$ : the variance of $\delta v_T$ as $\sigma_v^2$, $<\delta r_T>$ and $<\delta r_T>$ being their respective mean. In the "position" representation, the kick on position is a random Gaussian with

$$\delta r_T = \begin{cases} <\delta r_T> = 0 \\ \sigma_r^2 = 2D \, dt \end{cases} . \tag{9}$$

In the "velocity" representation, the kick on velocity is a random Gaussian variable $\delta v_T$ constructed so that, after time integration, it induces the same average mean dispersion on positions as $\delta r_T$ (i.e. $(\sigma_r^2)^{1/2} = dt \times (\sigma_v^2)^{1/2}$):

$$\delta v_T = \begin{cases} <\delta v_T> = 0 \\ \sigma_v^2 = \dfrac{2D}{dt} \end{cases}. \qquad (10)$$

Both methods are possible: one of the two representations must be chosen.

These kinds of procedures are known to yield good results for the random walk of a solute particle (i.e. dust) with constant $D_d$ inside a solvent (i.e. gas) of uniform density. It has been used several times in environmental studies, notably to study the diffusion of pollutants in the atmosphere or ocean (see e.g., Wilson & Sawford 1996). For transport of dust in the protoplanetary disk, Youdin and Lithwick (2007) and Hughes and Armitage (2010) use similar procedures in which $\delta v_T$ is a discrete random variable allowed to take 2 values $+(2D/dt)^{1/2}$ or $-(2D/dt)^{1/2}$.

However, problems arise when the solvent (i.e. gas) has a non-uniform density in space. In this case, the simple procedure described above is no longer valid. It can be easily verified that this method does not satisfy the "good mixing condition": at steady-state, a solute (i.e. dust) has a uniform concentration throughout the solvent (i.e. gas) in order to reach maximum entropy. In other words, it must have the same spatial density as the solvent times a constant multiplicative factor. The procedure described above (Eq. [9] or Eq. [10]) will lead inexorably to a uniform density distribution of the solute (i.e. dust) throughout the whole space, even though the solvent (i.e. gas) has a non-uniform density. For our astrophysical case, this means that dust would diffuse everywhere in space, out of the gaseous disk itself, in the absence of central star gravity. This problem has long been identified in environmental studies and several solutions exist with different derivations (see e.g., Sothl & Thomson 1999 or Ermak & Nasstrom 2000). Yet most of them are either expressed in terms of gas velocity fluctuations (which are not known or not "well documented" in the protoplanetary disk literature) or deal only with the self-diffusion of non-inertial material rather than being expressed in terms of diffusion coefficient, as is required here. For the case of protoplanetary disks, we wish to use only the diffusion coefficient and gas macroscopic properties in order to compare our results with previous analytical studies. To cure this problem of "good mixing", it could be tempting to introduce a spatial dependence in the diffusion coefficient $D$ to enforce a uniform concentration at steady-state. But this would run counter to the usual evaluations of the dust diffusion coefficient $D_d$ in a turbulent disk: in an $\alpha$-turbulent isothermal disk, $D_d$ is assumed to be close to the turbulent viscosity $D_d \sim \alpha\, C_s\, H$. Since $C_s$ and $H$ depend only on $r$ in an isothermal disk, $D_d$ is a constant function of $Z$. Hughes and Armitage (2010) identified this problem of dust-to-gas ratio and propose an empirical method in which $\delta v_T$ is a non-Gaussian discrete random variable. It is designed so that there is higher probability of a dust particle migrating toward higher density regions with weights depending on the local density profile. However, although this procedure is successful in the framework of their study, it is performed in the radial direction only and its extension to 3D systems is somewhat unclear since the number of possible directions becomes infinite. Ciesla (2010) presents a physically motivated derivation of the dust displacement algorithm, but treats the problem in the vertical direction only, for a disk density in the form $\rho(z) = \rho_0 \exp(-z^2/2H^2)$, and within the limit of very small particles that are strongly coupled to the gas and thus follow the same path as the gas molecules in the absence of turbulence. We present below a heuristic derivation to properly account for gas density variations. While sharing some similarities with Ciesla (2010), this derivation is more general and not constrained by all the previously mentioned limitations or assumptions.

**2.3 Good diffusion with a constant diffusion coefficient.**

For the sake of clarity, we restrict ourselves to the case of a constant diffusion coefficient in the current section. The more general case of lagrangian diffusion with a varying diffusion coefficient is treated in section 2.4. We first recall some basic principles of a 1D Gaussian random walk of a solute particle in a uniform and static solvent. We call X the position of a Lagrangian particle. For a particle with lagrangian velocity V and constant diffusion coefficient D, the spatial position increment dX due to random walk as described in Eq. (9) is:

$$dX = V\, dt + (2Ddt)^{1/2}\, W, \qquad (11)$$

where W is a Gaussian random variable with mean 0 and variance $\sigma^2=1$ (such that $(2Ddt)^{1/2} \times W$ is a random variable with mean 0 and variance $2Ddt$, as in Eq. [9]). A basic result of Einstein-Brownian diffusion is that the resulting motion has an average position $<X>$ so that $d<X>/dt=V$, and that the standard deviation $<(X-<X>)^2>$ evolves according to $d/dt(<(X-<X>)^2>)=2D$. In a Lagrangian simulation involving a large number of test-particles with positions that evolve according to Eq. (11), the local ensemble average is a solution to the Eulerian advection diffusion equation of a solute in a solvent with uniform volume density, which reads in Cartesian coordinates:

$$\frac{\partial \rho}{\partial t} + \frac{\partial \rho <V>}{\partial x} = \frac{\partial}{\partial x}\left( D\, \frac{\partial \rho_d}{\partial x} \right), \qquad (12)$$

also written as:

$$\frac{\partial \rho}{\partial t} + \frac{\partial}{\partial x}\left( \rho <V> - D\frac{\partial \rho}{\partial x} \right) = 0. \qquad (13)$$

Bearing in mind that the mass conservation equation reads $\partial\rho/\partial t + \partial F/\partial x = 0$, where F is the flux, the term $\rho<V>$ in the parenthesis of Eq. (13) is the advective flux and $-D\partial\rho/\partial x$ is the diffusive flux. Unfortunately, since the solvent (gas) density is not uniform, the transport equation of the solute (dust) does not have exactly the same form as Eq. (12). Thus, it cannot be solved using the simple method of Eq. (11) (equivalent to Eq. (9)). In a gaseous protoplanetary disk, the transport equation of dust in the gas disk is given rather by (see e.g., Dubrule et al., 1995, Takeuchi and Lin 2002):

$$\frac{\partial \rho_d}{\partial t} + \frac{\partial}{\partial x}\left( \rho_d V_d - \rho_g D_d \frac{\partial}{\partial x}\left(\frac{\rho_d}{\rho_g}\right) \right) = 0, \qquad (14)$$

where $\rho_g$ is the density of gas, $V_d$, $D_d$ and $\rho_d$ are the Eulerian velocity, the diffusion coefficient and the density of dust in the disk, respectively. The difference between Eq. (13) and Eq. (14) is that the term $D\partial\rho/\partial x$ is replaced by $\rho_g D_d \partial/\partial x\, (\rho_d/\rho_g)$, which accounts for gas density variations. Note that when $\rho_g$ is constant Eq. (14) reduces to Eq. (13), as expected.

To build a lagrangian approach, we wish to rewrite Eq. (14) into a form functionally close to Eq. (13), in which the diffusion term depends only on $\partial\rho_d/\partial x$. This is simply done by developing the diffusion term of Eq. (14):

$$\frac{\partial \rho_d}{\partial t} + \frac{\partial}{\partial x}\left( \rho_d \left( V_d + \frac{D_d}{\rho_g}\frac{\partial \rho_g}{\partial x} \right) - D_d \frac{\partial \rho_d}{\partial x} \right) = 0. \qquad (15)$$

By identifying Eq. (15) with Eq. (13) we can recover Eq. (13) simply by adding a correction term to the mean velocity of particles:

$$<V> = V_d + \frac{D_d}{\rho_g}\frac{\partial \rho_g}{\partial x} . \qquad (16)$$

This means simply that a gradient in gas density induces a net diffusive flux D/$\rho_g$×***grad***($\rho_g$) directed toward higher density regions. This correction term is a systematic component of the diffusive term arising from non-homogeneous diffusion (see e.g., Stohl & Thomson 1999; Ermak & Nasstrom 2000, Van Millgen et al. 2005). This is physically required in order to have a well-mixed final steady state.

We now come back to the mathematical implementation of dust diffusion in our lagrangian code, including correction for the gas density gradient. In the "position" representation, the kick on position is a random Gaussian variable $\delta r_T$ with mean < $\delta r_T$> and variance $\sigma_r^2$ now given by:

$$\delta r_T = \begin{cases} <\delta r_T> = \dfrac{D_d}{\rho_g}\dfrac{\partial \rho_g}{\partial x} dt \\ \sigma_r^2 = 2 D_d dt \end{cases} . \qquad (17)$$

So the displacement during one time-step is $\delta r_T$=< $\delta r_T$>+ W $\sigma_r$ (W being a normal random variable). In the "velocity" representation, the kick on velocity is a random Gaussian variable $\delta v_T$ with mean < $\delta v_T$> and variance $\sigma_v^2$ linked to $\delta r_T$ through the relation $\delta r_T$= $\delta v_T$×dt, so < $\delta v_T$>=< $\delta r_T$>/dt+W $\sigma_r$/dt or in other words:

$$\delta v_T = \begin{cases} <\delta v_T> = \dfrac{D_d}{\rho_g}\dfrac{\partial \rho_g}{\partial x} \\ \sigma_v^2 = \dfrac{2 D_d}{dt} \end{cases} . \qquad (18)$$

In the velocity representation, we first integrate the motion considering only the gas drag with the mean flow and with the star's gravity. At the end of the time-step, the particle's velocity is modified by adding the velocity kick as defined in Eq. (18).

To illustrate the validity of these results, we show in Figure 1 simple tests of 1D diffusion of dust in a non-uniform gas medium and with periodic boundary conditions. Gas density is given a sinusoidal density profile and all test particles are released at a same starting location, X=0.5. As there is no net transport, particles are subject to turbulent diffusion only. The diffusion coefficient is arbitrarily set to 0.001. In a first set of runs (Fig.1, left column), the density correction term is neglected such that $\delta r_T$ (or $\delta v_T$) for each particle is drawn from a random distribution according to Eq. (9) (kicks on positions). In Figure 1, we see that after 1,000 time-steps a steady state is reached in which the absolute spatial density of particles is uniform (Fig.1.a), whereas the gas is not uniform. As a result, the final dust concentration profile is not uniform (Fig.1.c), which is not physical.

In a second run, we include the correction term, thus following Eq. (17). The time evolution of the system is shown in the right column of Figure 1. We see that the system tends toward a state where dust concentration in the gas is asymptotically uniform (Fig.1b) and takes on a volume density profile similar to the profile of the gas, up to a constant multiplicative factor (Fig.1d), which is physical. These results illustrate that our numerical procedure naturally allows diffusion to tend towards a state of uniform concentration, in agreement with the "good mixing condition". This will be tested further in "real conditions" in Section 3.

**2.4 Good diffusion with a varying diffusion coefficient.**

We now consider a coefficient of diffusion that varies in space, so D is now written D(x). Such a behavior may be encountered in many physical situations : for example numerical simulations (see e.g. Fromang & Nelson 2009, Turner et al., 2010) shows that MRI turbulence is not uniform vertically in the disk and that the velocity fluctuations $\delta V_z$ tend to increase with Z. In consequence the effective gas diffusion coefficient increases according to $D_g \sim \delta V_z^2/\Omega$. This may be especially important in the case a "dead zone" (a laminar region) is present in the disk's midplane under an active upper layer. In this case $D_g$ may vary by several orders of magnitudes on distances of a few scale heights only.

It can be shown that when D(x) has a non-zero gradient, then the classic procedure of a random walk which step is a random variable with 0 mean and 2Ddt standard-deviation (like in Eq. 11) is not anymore an accurate solution to the equation $\partial X/\partial t = \partial^2 Dx/\partial t^2$. The right random variable to consider in this case has been studied by our colleagues in environmental studies (See e.g. Ermak & Nasstrom 2000), and is given by (see Ermak & Nasstrom 2000 and Annex 1) in the position representation :

$$\delta r_T = \begin{cases} <\delta r_T> = \dfrac{\partial D(x)}{\partial x} dt \\ \sigma_r^2 = 2D(x)dt + \left(\dfrac{\partial D(x)}{\partial x} dt\right)^2 \end{cases} \quad (19)$$

and in the velocity representation

$$\delta v_T = \begin{cases} <\delta v_T> = \dfrac{\partial D(x)}{\partial x} \\ \sigma_v^2 = \dfrac{2D(x)}{dt} + \left(\dfrac{\partial D(x)}{\partial x}\right)^2 \end{cases} \quad (20)$$

Ermak & Nasstrom (2000) suggests to increase the order of the random-variable and to consider a distribution with non-zero skewness (the skewness is the 3$^{rd}$ moment of a distribution, and is 0 for a symmetric distribution) to better the result. In the current paper we found excellent result considering only symmetric distributions (i.e a gaussian) as described by Eq.(19) or Eq.(20). We now need to put all things together to treat the most general case.

**2.5 Good diffusion : putting all things together**

We now treat the most general case of dust in the protoplanetary disk, with a varying dust diffusion coefficient $D_d$ and also with a varying gas density (the solvent). The good random variable for the dust random walk of is simply obtained by considering that in the absence of a gas-density gradient the random walk is described by Eq.(19) and that when a gas-density gradient is present, an additional term appears on $<\delta r_t>$ only given by Eq.(17). So in the position representation, $\delta r_t$ is now :

$$\delta r_T = \begin{cases} <\delta r_T> = \dfrac{D_d}{\rho_g} \dfrac{\partial \rho_g}{\partial x} dt + \dfrac{\partial D_d(x)}{\partial x} dt \\ \sigma_r^2 = 2D_d(x)dt + \left(\dfrac{\partial D_d(x)}{\partial x} dt\right)^2 \end{cases} \quad (21)$$

And in the velocity representation:

$$\delta v_T = \begin{cases} <\delta v_T> = \dfrac{D_d}{\rho_g}\dfrac{\partial \rho_g}{\partial x} + \dfrac{\partial D_d(x)}{\partial x} \\ \sigma_v^2 = \dfrac{2D_d(x)}{dt} + \left(\dfrac{\partial D_d(x)}{\partial x}\right)^2 \end{cases} \qquad (22)$$

Equations 21 and 22 are the core result of the present paper.

**2.6 Numerical considerations: kicks on position or on velocity?**

We have seen above that two methods are possible: either a kick on positions or a kick on velocities. For example, Ciesla (2010) uses a kick on position, whereas Hughes and Armitage (2010) and Youdin and Lithwick (2007) use a kick on velocities. Which is the better choice? Both methods have their own caveats and suffer from the fact that Brownian motion is still a crude physical model that hides unphysical infinite velocities and/or accelerations.

- A kick on position is an instantaneous transport that depends on the time-step (Eq. [21]). It induces a discrepancy between the velocity field and the position field because of the sheared nature of a Keplerian disk. This may, however, be partially cured by applying the kick at the intermediate time-step and by self-consistently correcting velocity at the end of the time-step.
- A kick on velocity depends on the time-step (Eq. [22]), inducing again velocity dispersion among particles that are time-step dependant.

After testing both procedures, it turns out that a kick on positions seem to yield somewhat better results for two reasons:

(a) in the limit of dt→0, which is often desirable for numerical accuracy and stability, the magnitude of kicks on positions converges toward 0 (Eq. [21]), which is numerically stable, whereas kicks on velocity diverge (Eq. [22]), which is numerically unstable.
(b) Since the velocity of small particles is efficiently damped by gas drag, velocity kicks may induce a strong gas drag that brakes accelerations $a_b$, the magnitude of which is $\sim\delta v_T/\tau \sim (2D_d/\tau^2\, dt)^{1/2}$. Generally, the time-step, dt, is a small fraction of the orbital period i.e. dt=$\varepsilon/\Omega_k$ with $\varepsilon\sim$0.1% to 10%. Since $D_d$ is of the order of the turbulent viscosity $D_d\sim\alpha\, H^2\Omega_k$, $a_b$ is proportional to $\Omega_k/(\tau\varepsilon^{1/2})$. Small particles with a very small Stokes number will thus be subject to very high accelerations requiring very small integration time-steps to ensure accuracy and/or stability. Reducing the time-step will induce a decrease of $\varepsilon$ and will not solve the problem as it increases the velocity kick. It may then be impossible to integrate very small particles in the limit St→0.

For these reasons, integration with velocity kicks appeared to be about five times slower than with position kicks (for particles with St>$10^{-5}$) using an implicit Bulirsch-Stoer solver with adaptive time-step control (Press et al. 1992). We thus chose to integrate particle motion using kicks on positions. This yields good results for all particle sizes, as shown in Section 3.

**2.6 Expressions of the diffusion coefficient**

Various expressions exist in the literature for $D_d$. In a turbulent protoplanetary disk it is generally thought that turbulence is the main process driving the outward transport of angular momentum (and thus mass accretion) and diffusion of dust. Numerical simulations (see e.g., Fromang & Nelson 2009) have shown that the effective gas diffusion coefficient, $D_g$, is close to the turbulent viscosity, that describes the transport of angular momentum. The diffusion coefficient $D_d$ is linked to the turbulent viscosity through the Schmidt number Sc, defined as the ratio of the gas-diffusion coefficient ($D_g$) to the dust diffusion coefficient ($D_d$). In other words, $Sc=D_g/D_d$. $D_g$ is approximately $\alpha C_s H$ as measured in numerical simulations of MRI turbulence (where $\alpha$ is the turbulent viscosity parameter and H is the local pressure scale height of the gas disk; see e.g., Fromang & Papaloizou 2006) this gives:

$$D_d \approx \frac{\alpha C_s H}{Sc}. \qquad (23)$$

Sc depends on the particle size. It tends to 1 in the limit of very small particles and increases to infinity for very big particles. This means that very big particles no longer "feel" the turbulence because of their inertia. The expression of Sc is a matter of debate and, in the literature, depends on the value given to St. Following the work of Youdin and Lithwick (2007) we assume the correlation time to be the eddy time ($\tau_c = \tau_{edd}$). The eddy turnover time is $\tau_{edd} \sim 1/(\alpha^{2\gamma-1} \Omega_k)$. The value of $\gamma$ depends on the structure of the turbulence: if turbulent diffusion is driven by large slow-moving turbulent eddies then $\gamma \rightarrow 1$, but if driven by small fast-moving turbulent eddies $\gamma \rightarrow 0$ (Brauer *et al.* 2008). However, several studies (Cuzzi et al. 2001; Schräpler & Henning 2004) report $\gamma=1/2$, such that $\tau_{edd} \sim 1/\Omega_k$ ($\Omega_k$ representing the local Keplerian frequency) in agreement with Fromang and Papaloizou (2006), who measure $\tau_c \sim 0.15$ orbital period. In LIDT3D, the time correlation in the kicks was introduced stochastically, similarly to Youdin and Lithwick (2007): at each time-step and for every particle a random number is generated and the probability of applying a kick is the ratio $dt/\tau_c$. Thus, particles close to the central star will be subject to kicks much more frequently than those at greater distances. This is illustrated in Section 4.2.

The expression of the Schmidt number is a matter of some debate. Dullemond and Dominik (2004), for example, use Sc=1+St. Recently, however, in their paper on diffusion driven by anisotropic turbulence, Youdin and Lithwick (2007) suggest that $Sc=(1+\Omega_k^2\tau^2)^2/(1+4\Omega_k\tau)$, for radial diffusion only, on the basis of an analytical study coupling sedimentation and orbital oscillation. This is in agreement with the numerical simulations of Carballido et al. (2006). This expression will be used in our code unless otherwise specified. In addition to these analytical expressions, some recent papers (Fromang & Papaloizou 2006; Fromang & Nelson 2009) have attempted to characterize the diffusion of small dust particles using direct 3D MHD simulations. They show that the vertical distribution of dust does not fit a Gaussian distribution (as often assumed in analytical models), with major discrepancies for the upper layers (Z/H>1). As explained by the authors, this may be due to larger velocity fluctuation ($\Delta V_z$) of the gas for increasing values of Z/H. As a result, D increases quadratically with Z and has a value of $\sim 0.002 C_s H$ at the midplane. They provide a basic fit to their results in Eq. 26 of Fromang and Nelson (2009). Non-uniform dust diffusion will be presented in Section 4.1.

**3. Testing the turbulence diffusion algorithm**
In this section, we present the different tests used to benchmark the diffusion algorithm against either analytical or numerical models in some simple cases.
**3.1 Gas disk model**
In the rest of this paper, a 3D analytical model of a locally isothermal gas disk is used for the sake of simplicity. It provides gas density, temperature and velocity field. However, the method presented in Section 2 is easily applied to any disk sampled on a 3D grid provided that the density gradient can be computed. Here we use the disk structure described in Takeuchi and Lin (2002). The 3D gas density as a function of R and Z is:

$$\rho_g(r,z) = \rho_0 r^p e^{\frac{-z^2}{2H(r)^2}} \qquad (24)$$

where $H^2(r)=H_0^2 r^{q+3}$ and $\rho_0$ is the gas density at unit radius, and p and q are two constants. With these notations, the exponent of the surface density profiles is $s=p+(q+3)/2$ (Takeuchi & Lin 2002). We use the fiducial disk model of Brauer et al. (2008) with indexes $p=-2.25$ and $q=-0.5$, corresponding to a surface density with exponent $s=-1$, a gas surface density of 200 kg/m² and a temperature of 204K at 1 *AU* with a central star mass of $0.5 M_\odot$. Unless otherwise specified, the turbulent parameter is $\alpha=0.01$ and uniform throughout the disk. In order to isolate transport effects due to diffusion and sedimentation from those caused by advection in the gas flow, we fix the radial and vertical gas velocity at zero. Consequently, the nebula considered in the following tests is simply a modified minimum-mass nebula. In a forthcoming paper, this simple (but convenient) model will be replaced by a 2D time-evolving disk, as in Ciesla's work (2009).

**3.2 Vertical diffusion of particles: testing small and big particles**

Turbulent diffusion opposes dust settling in the vertical direction. The dust thus attains a vertical steady-state distribution, which in general is not a Gaussian (strictly speaking). For small dust particles that reach terminal velocity rapidly (in less than one orbital period), the equation governing the evolution of equilibrium dust density along the Z direction is (Dullemond & Dominik 2004; Dubrulle et al. 1993):

$$\frac{\partial \rho_d}{\partial t} = \frac{\partial}{\partial z}\left(\rho_d \tau \Omega_k^2 z\right) + \frac{\partial}{\partial z}\left(D_d \rho_g \frac{\partial}{\partial z}\left(\frac{\rho_d}{\rho_g}\right)\right) \qquad (25)$$

The term $\rho_d \tau \Omega_k^2$ is the particle's terminal velocity in the gas, which is assumed to be reached in a short time (in other words, the stopping timescale is much shorter than the diffusion and orbital timescales, i.e St<< 1), which means that Eq. (25) applies to small particles only. When $D_d$ is constant, this equation can be analytically solved to determine the steady-state vertical distribution of tightly coupled particles (Dubrulle et al. 1995; Fromang & Nelson 2009):

$$\rho_d = \rho_{d,mid} \exp\left(\frac{(\Omega_k \tau)_{mid} C_s H}{D_d}\left(e^{\frac{z^2}{2H^2}} - 1\right) - \frac{z^2}{2H^2}\right) \qquad (26)$$

For particles with stopping times much larger than their orbital period and which are loosely coupled to the gas ($\Leftrightarrow \Omega_k \tau \geq 1 \Leftrightarrow St\geq 1$ assuming that the eddy turnover time is $\sim 1/\Omega_k$), vertical motion is mainly driven by their orbital oscillation. The scale height $H_p$ of these big particles can be estimated using the argument of Youdin & Lithwick (2007). The time they need to return to the midplane after a vertical perturbation is given by their stopping time $\tau$. Thus, $H_p$ is easily found by equating the diffusion time over $H_p$ ($\sim H_p^2/D_d$) to the stopping time ($\tau$) implying $H_p=(D_d \tau)^{1/2}$ (equivalent to $H_p=(D_g/\tau\Omega_k^2)^{1/2}$ reported in Eq.6 of Youdin & Lithwick 2007 since $D_d \sim D_g/(\tau\Omega_k)^2$ according to their Eq.4). This is more naturally expressed in terms of the Stokes number ($\tau \sim St/\Omega_k$), which gives $H_p=(D_d St/\Omega_k)^{1/2}$. Note that $D_d$ depends on the particle's size through the Schmidt number. $H_p$ is equivalent to Eq. (6) of Youdin & Lithwick (2007) but, in our expression, $D_d$ is not explicitly given in terms of the particle stopping time. Assuming that particles are vertically distributed along a Gaussian with scale height $H_p$, it is possible to derive a simple expression for the vertical distribution of particles in the regime St≥1 (loose coupling):

$$\rho_d = \rho_{d,mid} e^{-\frac{z^2 \Omega_k}{2(D_d \cdot St)}} \qquad (27)$$

We now verify below whether the analytical distributions of dust in both the tightly coupled regime (Eq. [26]) and loosely coupled regime (Eq. [27]) are reproduced with the method described in Section 2.5, which is the core of the LIDT3D code.

We first run a simulation with particles orbiting in an annulus extending from $r_{min}$=1 AU to $r_{max}$=1.01 AU in a gaseous disk as described in Section 3.1. We use periodic radial boundary conditions (i.e. when a particle is found at position $r_{max}$+ dr , dr>0, it is shifted to position $r_{min}$+dr modulo $r_{max}$-$r_{min}$). The particle's velocity is also corrected to account for the keplerian shear. Initially 70,000 different-sized particles (10,000 particles per size bin: 0.1, 1, 10, ..., $10^5$ microns) are uniformly distributed vertically from Z = -2H to Z = +2H. Particles evolve under the action of gas drag and turbulent diffusion, using the algorithm described in Section 2.5 (note here that $D_d$ is constant with height so that all terms in $\partial D_d/\partial Z$ vanish). The Schmidt number is assumed to be a constant, Sc=1.5. In Figure 2, particle location is plotted after 20,000 orbits and clearly shows that particles are vertically sorted according to their sizes, with the smallest dust particles floating in the upper layers and the largest orbiting close to the midplane.

Vertical distributions of same-sized particles are plotted in Figure 3 (solid line) and compared to Eq. (26) (dashed line) and Eq. (27) (dotted line in Fig.3). The vertical distribution obtained with LIDT3D closely matches the analytical estimates as showed hereafter. We see in Figure 3 that, for $\Omega_k\tau$ ranging from $10^{-6}$ to $10^{-2}$, LIDT3D produces an excellent solution, even though the integration time-step is about $10^4$ times the particle stopping time owing to the efficiency of the implicit solver. For $\Omega_k\tau$> 0.1, the assumption of tight gas-particle coupling breaks down. This explains the discrepancy between the dashed and solid lines in Figure 3a. To explore the loose-coupling regime, an additional simulation is run for particles with $\Omega_k\tau$=2.5 and 25, with the assumption that the Schmidt number increases with the particle stopping time Sc~1+$(\Omega_k\tau)^2$ (valid for radial diffusion, as stated in Youdin & Litchwick 2007, whereas it is used here as an isotropic diffusion coefficient, as in Birnstiel et al, 2010). The result is plotted in Figure 4 against the analytical models for tight coupling (Eq. [26]) and loose coupling (Eq. [27]). The distributions produced by LIDT3D are closely matched by the analytical distribution for loose coupling (dotted line).

These tests show that for both small and large particles, LIDT3D produces a good physical solution. An additional test is also presented in Annex 2 for the case of a vertically varying diffusion coefficient. Any particle size can thus be treated self-consistently with a single tool, from sub-micron dust to macroscopic boulders. However, note that for dust particles at St>>1, large-scale correlated motion may also arise in MRI simulations (i.e. velocity fluctuations with long correlation distances, S. Fromang, private communication). We have not yet taken these into consideration. However, as this is not an intrinsic limitation of the method, we plan to introduce such spatial correlations into the system in the near future.

**3.3 Radial diffusion in the midplane**

To test radial diffusion against simple models, we run a purely 2D simulation of dust transport in the midplane (Z=0). We consider only very small particles (0.1 micron), with Stokes numbers in the range $10^{-7}$ to $10^{-8}$ between 1 and 10 AU. Since the particles are tightly coupled to the gas, which is not moving radially (Vr = 0), their radial motion results uniquely from turbulent diffusion (with α=0.01). 10,000 particles were released at 10 AU and evolved using the procedure described in Section 2.5.

The resulting particle volume density as a function of distance is reported in Figure 5 (colored lines). As expected, the density distribution widens with time, adopting a somewhat asymmetric shape owing to the tendency for particles to mix well with the gas. We compare the distributions obtained with our stochastic algorithm with the numerical integration of the advection-diffusion equation in the radial direction given by:

$$\frac{dC}{dt} = \frac{1}{\rho_g r} \frac{\partial}{\partial r}\left(-rV_r \rho_g C + r\rho_g D_d \frac{\partial C}{\partial r}\right) \quad (28)$$

where C is the concentration of dust in the gas, $D_d$ is the diffusion coefficient and $V_r$ is the radial velocity of the gas (=0 here). Eq. (28) is numerically integrated using a second-order derivative operator (centered difference) and a first-order time-scheme with a time-step of about 0.1 year. The result from this Eulerian model is shown by the black solid line in Figure 5. The agreement with LIDT3D (colored lines) is excellent, apart from the regions below 2 AU at T>10,000 years due to edge effects (all particles passing below 1 AU are eliminated from our particle simulation in this example).

In the examples considered in the present paper (section 3 and 4) numerous testes showed that the $\partial D_d/\partial x$ appeared to play solely a minor role as far as $D_d$ is defined as a constant fraction of $C_sH$ (like in the standard formulation $D_d \sim \alpha C_s H$ with α=cst) which is the standard case of the current paper. However, for disks containing a central "dead-zone" in their midplane, $D_d$ may be strongly varying with Z. Thus the gradient of $D_d$ may even dominate in some cases. These results will be presented in a forthcoming paper concentrating on the subject of diffusion in dead-zones.

To conclude this section, we have shown that our method for calculating Lagrangian diffusion fulfills the "good mixing" constraint for both vertical and radial diffusion; it is in close agreement with analytical models for vertical mixing (in regimes of both tight and loose coupling to the gas), and in excellent agreement with the numerical integration of the diffusion-advection equation for radial diffusion.

## 4. Applications

### 4.1 Comparing midplane and out-of-plane diffusion

In many studies, both analytical and numerical, dust diffusion is studied along the radial direction only, so as to make computation tractable. In order to do this, authors solve the radial transport equation (Eq. [28]) by applying one of the following three assumptions: either they average dust density in the vertical direction and assume perfect mixing (Gail 2001), or consider diffusion only at the disk's midplane (Hughes and Armitage 2010), or vertically integrate the diffusion equation assuming a Gaussian vertical distribution of dust (Brauer et al. 2008; Birnstiel et al. 2010). Moreover, in most of the aforementioned studies, particle dynamics is cast in the simplified form of asymptotic dynamics (at times much larger than the stopping time), and is most often computed at the midplane only. Whereas these approaches are very useful for understanding the large-scale properties of dust diffusion, the fact that they do not explicitly resolve the Z direction may lead to uncertainties or biases. None of the aforementioned approximations are really physically justified as, indeed, dust particles never reach perfect vertical mixing with the gas (see Section 3.2) or, due to turbulence affects, remain confined to the midplane. They are always in a somewhat intermediate state. At this point, we would emphasize the need to track dust motion in the R and Z directions simultaneously and provide some comparisons of three-dimensional simulations against purely midplane cases. It is to be expected that dust diffusion is more effective in the disk's upper layers for the following reasons: for a gas disk structure such as in Eq. (24), the magnitude of the gas density gradient is smaller in the disk's upper layers than at the midplane (see top of Fig.6). This has two important consequences:

(a) As described in Takeuchi and Lin (2002), for Z>1.5 H the gas rotation profiles become super-Keplerian, inducing an *outward* gas drag force (see Section 3.1 of Takeushi & Lin 2002), which favors an outward migration of dust particles. This effect cannot be captured unless the Z direction is explicitly taken into account, as is the case here (see below).

(b) As dust particles need to reach a perfect mixing with gas in a Lagrangian description, they are subject to a diffusion velocity that is proportional to $D_d/\rho_g \times$ **grad**$(\rho_g)$, which is strongly directed inward for Z=0. However, as Z increases and the density gradient decreases, this inward diffusion velocity becomes increasingly weaker (in magnitude) and can even become positive for some values of Z/R (see the example in Fig.6 at bottom). The net effect is again an increased efficiency of diffusion in the disk's upper layers.

All these elements favor a non-homogeneous diffusion in the vertical direction, in which diffusion efficiency is an increasing function of altitude. We present below some illustrations of this, comparing them to the case of dust dynamics in the midplane only. It is far beyond the scope of the current paper to quantify these effects precisely, but we wish to provide examples in a disk with a surface density decreasing as $r^{-1}$ and with $\alpha$ = 0.01.

As a qualitative example, we first run a simulation using 5,000 particles, with sizes ranging from 0.1 micron to 1cm, starting at 10 AU. We prevent radial migration for 500 years so as to reach a vertical steady state and then allow the system to evolve radially (Fig.7). The Schmidt number here is Sc=$(1+\Omega_k^2\tau^2)^2/(1+4\ \Omega_k\tau)$ (valid for diffusion in the radial direction, Youdin & Litchwick 2007). As expected there is a clear gradient of particles with Z. The particle cloud does not spread symmetrically around the release location. In fact, particles with Z>H spread radially more rapidly (Fig. 7 at 3,000 and 5,500 years) than those with Z<H.

To observe the spatial and time evolution of dust distribution and to emphasize the differences with the case of midplane diffusion only, we run two simulations with 0.1-micron particles: one fully three-dimensional (3D cases) and the other two-dimensional confined to the disk's midplane (2D case). In both cases, 50,000 particles are initially released at 5 AU from the central star. The resulting surface density of dust as a function of time is plotted in Figure 8. After a time, particles spread in the disk and try to reach a "good mix" with the gas disk, and thus rapidly adopt a power-law surface density profile decreasing with R. The difference between the 2D and 3D cases is striking: in the 3D case, particles spread outwards much more rapidly than in the 2D case, In addition, the particle surface density is systematically higher in the 3D case than in the 2D case for regions further than the initial release point. This result is easily explained: in the 2D case, due to the strong negative density gradient of gas in the midplane, the diffusive flux is strongly directed inwards due to the diffusive flux in $D/\rho_g \times$grad$(\rho_g)$, in such a way that outward diffusion is to a certain extent prevented. Conversely, in the 3D case, particles can explore the upper layers of the disk where the gas density gradient is much shallower, or even positive, which thus favors outward diffusion. Note that a zero radial velocity of the gas is used in order to isolate the diffusion effect.

The present example illustrate simply that dust transport may be very significantly different when we consider explicitly the vertical direction of the disk, and that simulation of dust diffusion in the midplane only, or when perfect vertical mixing is assumed, may lead to large uncertainties, or even errors.

**4.2 Tracking particle paths in protoplanetary disks: a bridge toward cosmochemistry**

Our second example involves the direct tracking of dust motion in protoplanetary disks. One of the strengths of a Lagrangian description is that individual particle paths in the disk can be extracted from simulations. This can be useful for deciphering the thermochemical histories of dust and grains that will be incorporated into meteorites or planets. In Figure 9 to Figure 11, we present different examples of the trajectories of particles with increasing sizes. Initially, all the particles were released in the disk's midplane at 5 AU from the central star. Different behaviors can be readily identified: on the one hand, small micrometer-sized particles have very stochastic paths in the R/Z plane, with their (R,Z) location increasing or decreasing stochastically (see e.g., Fig.9 and Fig.10). At larger distances, we also notice that the frequency of turbulent kicks decreases while their magnitude increases: this is due to the eddy correlation time (see Section 3.3), which increases with the distance to the star. As a result, turbulent kicks are less frequent for increasing values of R, while the magnitude of the kick (the diffusion coefficient), such as the turbulent viscosity, increases with R ($\propto R^{q+3/2}$ and q=-0.5 represents the radial exponent index of the gas disk temperature). Our last example involves the dynamics of a 10 cm radius particle (Fig.11), which shows significant differences with the previous cases, as it drifts rapidly toward the central star due to strong gas drag. As is clearly visible in Figure 11, it is also subject to vertical oscillations around the disk's midplane, the amplitude of which diminish as the particle gets closer to the star.

During their journey in the disk, the dust particles go through different episodes of heating or cooling (see the bottom of Fig.9 to Fig.11) which may affect their chemical and isotopical composition due to multiple adsorptions and desorptions of the surrounding gas. Similarly, refractory inclusions in meteorites and comets have undergone multiple heating events in gaseous environments with drastically different oxygen isotopic compositions ($^{16}$O-enriched solar composition and $^{16}$O-depleted planetary composition) and oxygen fugacity (reducing and oxidizing). Thus tracking their trajectories together with their thermodynamical histories may reveal critical clues on how oxygen reservoirs were distributed in the solar protoplanetary disk. Because oxygen is the major rock-forming element, understanding how its isotopic composition evolved from solar to planetary is key to understand the physics of planetary formation.

These trajectories will be used in a forthcoming paper to study the chemical evolution of dust during its transport in the protoplanetary disk.

**5. Conclusion**

In this paper, we have presented a numerical tool (LIDT3D) for simulating three-dimensional transport of dust in a turbulent gaseous protoplanetary disk using an implicit Lagrangian approach, in order to allow tracking of individual particles in gas. The gas disk is treated separately from the dust and no retro-action of dust on gas is considered for the moment. We provide a simple formalism (Section 2) in which dust motion is separated into two parts: a deterministic motion due to the interaction with the mean velocity field of the gas and a stochastic part in order to account for the kicks induced by the turbulent gas velocity field. The construction of a good random variable compatible with diffusion coefficients and respectful of the Second Law of Thermodynamics is the core result of the paper (summarized in Eq. (21) and Eq. (22)). In particular, we have shown that in a Lagrangian form, diffusion induces a systematic velocity component proportional to $D_d/\rho_g\mathbf{grad}(\rho_g)$, which points along the direction of the local gas density gradient and satisfies the asymptotic good mixing condition. This systematic velocity term favors strong inward diffusion of dust in the disk's midplane, whereas in the disk's upper layers much more moderate inward diffusion occurs. Under some conditions, a net outward flux is even observed. Moreover, If the diffusion coefficient depends on r, an flux appears proportional to **grad**($D_d$) which direction depends on the specific expression for $D_d$. Thus it would seem necessary always consider diffusion in three-dimensions rather than using a 1D approach. We have presented different tests showing that, in both the vertical and radial directions, our LIDT3D tool accurately reproduces analytical models, when they exist, and well-known Eulerian 1D dust transport models. In the final section, we have described two initial applications. In the first example, we show that diffusive transport in the upper layers of the disk is more efficient than in the disk's midplane and may significantly increase the radial transport of dust. In particular, we show that the density of particles diffusing in the disk's midplane only is about 5-10 times less than for the 3D case, in which particles are allowed to move vertically in the disk (Fig.8). In a second set of examples, we present examples of individual particle paths for particles in the range 1 micron to 10 cm (Fig.9 to Fig.11). Among the cosmochemical implications of temperature variations is the amplitude of gas-solid interactions, such as adsorption-desorption of volatiles at low temperatures (~100 K) or condensation evaporation at high temperature (1000 K or more). Therefore, depending of the particle path, dust can be enriched or depleted in volatile or moderately volatile elements, and chemical as well as isotopic fractionations could occur due to interaction with the gas.

The next step will be to implement a time-dependent model of a gas-disk in LIDT3D, rather than a static one as presented here. We then plan to use this to compute the equilibrium distribution of dust in observed protoplanetary disks and couple the results with a radiative transfer code to compare them with observations. Another interesting aspect would be to study the effect of a dead-zone on particle dynamics and to see what level of dust density can be reached in these regions of low turbulence. Coagulation and fragmentation will be implemented using the formalism developed for debris disks in Charnoz and Morbidelli (2004, 2007).

An interesting application of our procedure is for SPH codes, which are Lagrangian in essence. As it is sometimes difficult to capture the details of turbulence with SPH codes, it could be useful to have a light and quantified procedure (as presented in Section 2) to introduce the diffusive effect of turbulence. In SPH simulations of gas and dust (as in Barrière-Fouchet et al. 2005 or in Fouchet et al. 2010m which includes a planet), the method presented in Section 2 provides a direct way to introduce a perturbation into dust-particle motion that mimics the effect of turbulence without the need for an explicit MHD code algorithm. This could be done provided that SPH noise could be reduced below the level of turbulent diffusion, and by extracting the laminar gas-velocity field. This approach will be tested in the near future. However one of the biggest uncertainties in the turbulent diffusion prescription is that ideal MHD does not apply in protoplanetary disks. As a result, accretion is likely to be layered (Gammie, 1996) with a non-turbulent dead zone of a few gas scale-heights near the midplane (see e.g., Fleming & Stone, 2003; Oishi & Mac Low , 2009; Turner & Carballido 2010). This will be studied in a future work using a spatially varying diffusion-coefficient.

For cosmochemical applications, LIDT3D will first be used to build samples of particle trajectories in the disk and to estimate their illumination and irradiation history, as well as their capacity to adsorb volatiles on their surface. The fact that this code can explicitly describe the instantaneous motion of dust in the disk and also provide the thermodynamical conditions of the surrounding gas experienced by a dust particle during its transport in the disk means that it can contribute greatly to bridging the gap between cosmochemistry (which constrains the composition of meteorites) and protoplanetary disks physics (constraining a disk's large-scale properties). Among the many cosmochemical applications possible, we will first focus on the transport and thermodynamical history of high-temperature components such as chondrules and CAIs, which constitute the majority of planet-forming materials. We will also study how this history links to the isotopic properties of their volatile constituents, such as noble gases and oxygen, which may transport isotopic anomalies implanted during their journey in the disk (Clayton et al. 1973; Lyons et al. 2009).

**ACKNOWLEDGEMENTS**

SC thanks Sebastien Fromang and Neal Turner for enlightening discussions, as well as all the members of the ANR DUSTYDISK group. We are indebted to an anonymous referee whose careful review, patience and tolerance resulted in a much improved paper. We thank also Gill Gladstone for her careful proofreading of the manuscript. This work was supported by the project DUSTYDISK of the French *Agence Nationale pour la Recherche* (ANR), with contract number ANR-BLAN-0221-07. It was also supported by the *Université Paris Diderot* with a CAMPUS SPATIAL grant for the project "*Formation et transport des premiers solides dans le Système Solaire*".

**ANNEX 1 :**

We describe how to derive the distribution of particles' positions from the equation of diffusion. We report a somewhat simplified version of the derivation presented by Ermak & Nasstrom (2000). Let P(x,t) the density distribution of the particles in space. If particles are in a solvent with a uniform spatial density, then their concentration C(x,t) is simply proportional to P(x,t). Assuming that the evolution of C(x,t) obeys to the diffusion equation, then P(x,t) is also a solution of the diffusion equation that reads :

$$\frac{\partial P}{\partial t} = \frac{\partial}{\partial x}\left(D(x)\frac{\partial P}{\partial x}\right) \quad (A1)$$

We assume that D is a function of x. From Eq.A1 it is possible to compute the successive moments of x. They are defined as

$$<x>^n = \int_{-\infty}^{+\infty} x^n P(x,t) dx \quad (A2)$$

So combining Eq.A1 and Eq.A2 we obtain

$$\frac{\partial <x>^n}{\partial t} = \int_{-\infty}^{+\infty} x^n \frac{\partial}{\partial x}\left(D(x)\frac{\partial P(x,t)}{\partial x}\right) dx \quad (A3)$$

Then we perform a 1$^{st}$ order Taylor expansion of D(x) around x=0: D(x)≈$D_0$+x$D'$. We assume also that P(x) and $x^n \partial P/\partial x$ converges towards 0 (at least for n=1 and 2) when x becomes infinite. Under these hypotheses, and after integration by parts of Eq.A3, we get the first moments (Ermak & Nasstrom 2000):

$$\frac{\partial <x>}{\partial t} = D' \quad (A4)$$

$$\frac{\partial <x>^2}{\partial t} = 2D_\circ + 4D'<x> \qquad (A5)$$

Equations A4 and A5 form a system of coupled differential equations. After integration from time= 0 to dt we obtain :

$$<x> = D'dt \qquad (A6)$$

$$<x>^2 = 2D_0 t + 2(D'dt)^2 \qquad (A7)$$

Since the standard deviation is $\sigma_x^2 = <(x-<x>)^2>$ we obtain :

$$<x> = D'dt \qquad (A8)$$

$$\sigma_x^2 = 2D_0 dt + (D'dt)^2 \qquad (A9)$$

We recover the common results that when D is constant (<=> D'=0 here) the mean of the displacement is 0 while the standard deviation increases like $2D_0 dt$, conversely when D'≠0 we recover the distribution reported in Eq. 19

**ANNEX 2:**

We detail the analytical derivation of the steady-state vertical distribution of dust when the diffusion coefficient depends on Z. The result is used to test the validity of the steady-state solution we obtain with our lagrangian code when $D_d$ is given by (Eq.29).

We start from the diffusion equation assuming that $\tau \ll 1/\Omega_k$ so that the particles vertical velocity is almost equal to their terminal velocity:

$$\frac{\partial \rho_d}{\partial t} = \frac{\partial}{\partial z}\left(\rho_d \tau \Omega_k^2 z\right) + \frac{\partial}{\partial z}\left(D_d(z)\rho_g \frac{\partial}{\partial z}\left(\frac{\rho_d}{\rho_g}\right)\right) \quad \text{(A10)}$$

After writing $d\rho_d/dt=0$, dropping the $\partial/\partial z$ term on both sides and dividing by $\rho_g$ we get:

$$\frac{\partial C}{\partial z} + \frac{C \tau \Omega_k^2 z}{D_d} = 0 \quad \text{(A11)}$$

where $C=\rho_d/\rho_g$ is the dust concentration. This is a first order differential equation. Its solution is:

$$C(z) = Exp\left(-\int_0^Z \frac{\tau \Omega_k^2 z}{D_d} dz\right) \quad \text{(A12)}$$

Also written as (by removing constant terms from the integrals and developing the expression of the stopping time):

$$C(z) = Exp\left(\frac{-\Omega_k^2 a \rho_s}{C_s^2}\int_0^Z \frac{z}{\rho_g(z)D_d(z)} dz\right) \quad \text{(A13)}$$

C(z) can be numerically computed for any expression of $\rho_g(z)$ and $D_d(z)$.

The dust density is then:

$$\rho_d(z) = C(z)\rho_g(z) \quad \text{(A14)}$$

Eq. (A13) and (A15) can be used to test the validity of our numerical approach by checking that our steady-state solution is the same as given by equations (A13) and (A14). We numerically compute the dust equilibrium distribution in the case $D_d(z)$ is a function of Z as in Eq.29 (see section 4.1). The result is shown in Figure A1. Our lagrangian method agrees very well with the analytical steady-state distribution.

# FIGURES

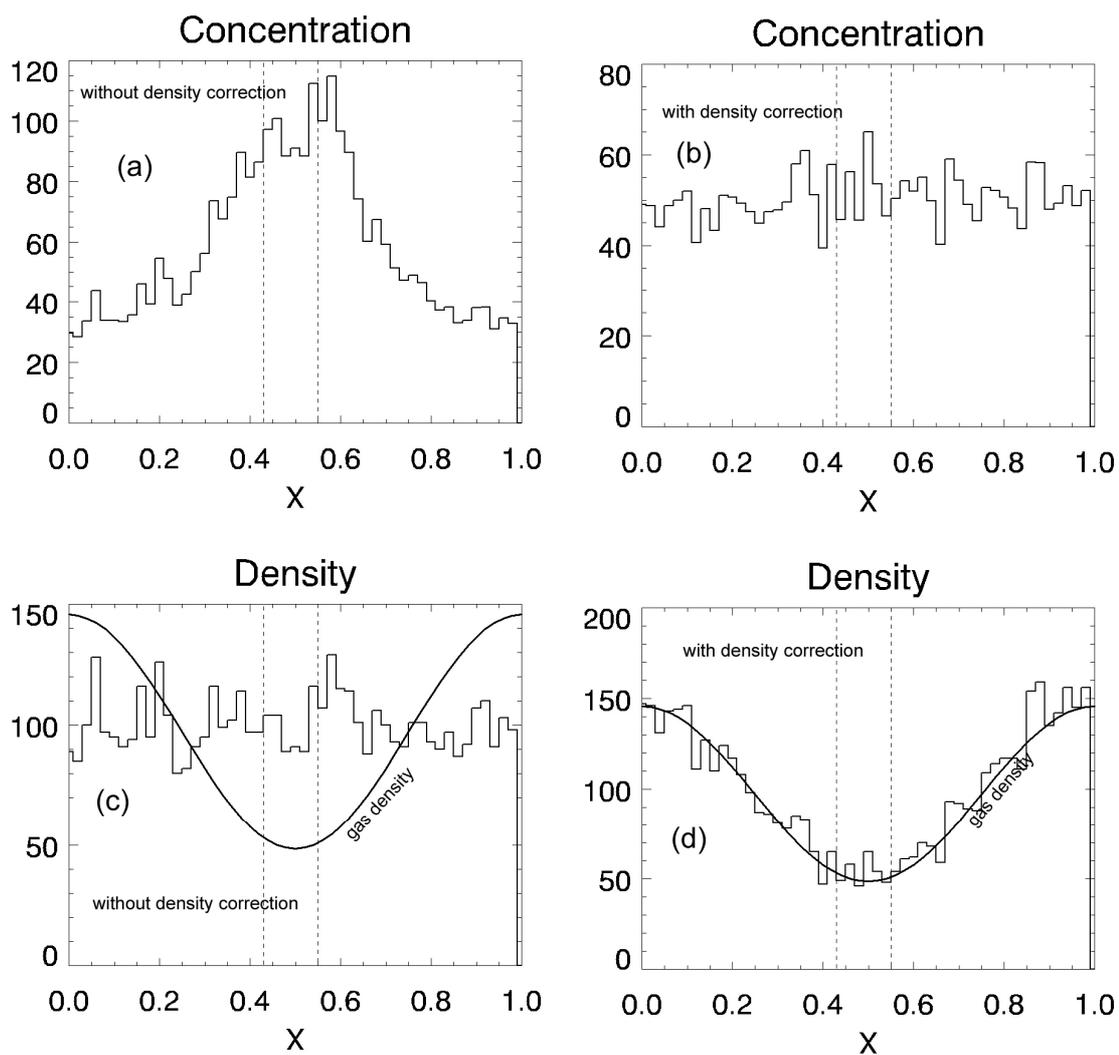

**Figure 1**: Test of 1D particle diffusion of a solute (particles) in a solvent (gas) with (b,d) and without (a,c) corrections for the density of the solvent. Top row: The solid line shows the number of particles in each X bin of the simulation; the dashed line shows the region where all particles were initially released. Bottom row: density of particles (number of particles in each X bin divided by the local gas density). The solid histogram shows the particles; the dashed line shows the initial state and the solid sinus curve shows the absolute density of the gas. Left column: without using the density corrections, right column: using the density correction.

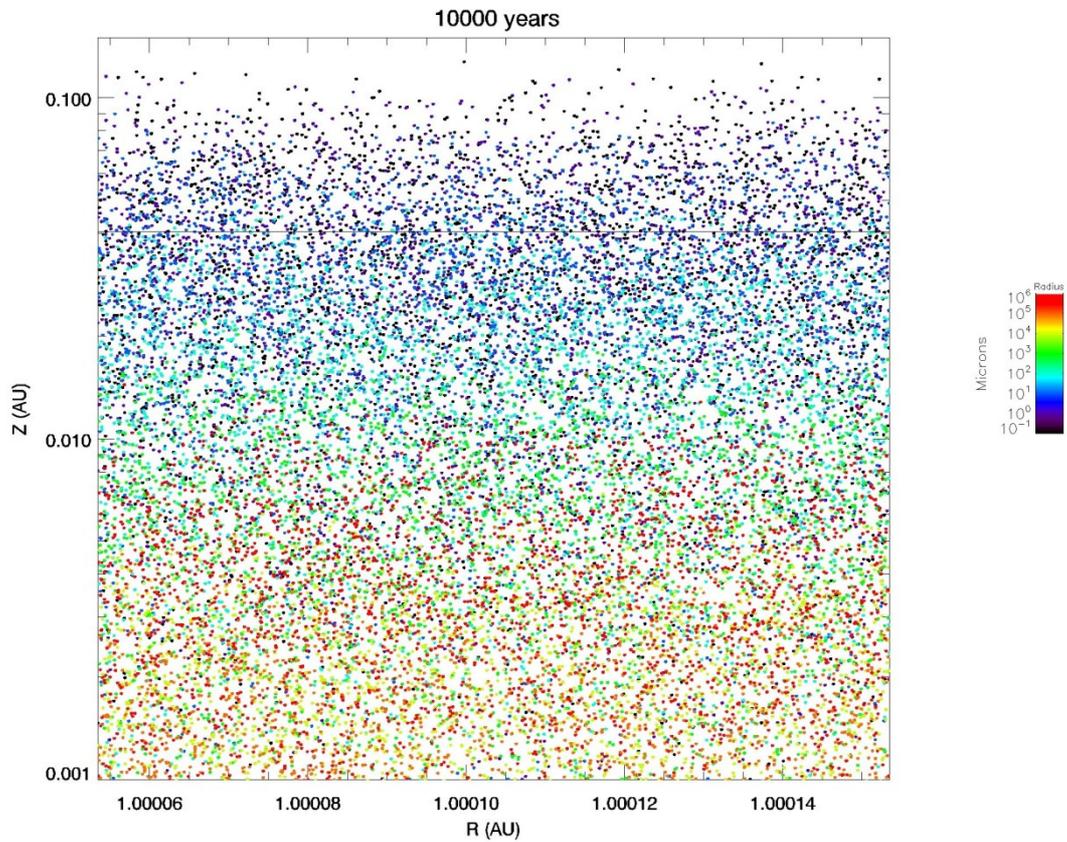

**Figure 2**: Location of 10,000 different-sized particles (color-to-size correspondence is given on the right scale) after 10,000 years evolution at 1 AU . A constant diffusion coefficient was used. We clearly see the sedimentation process: small particles are widely spread vertically, whereas big particles sediment rapidly and gather close to the midplane (z=0). The solid horizontal lines show the location of the pressure scale height.

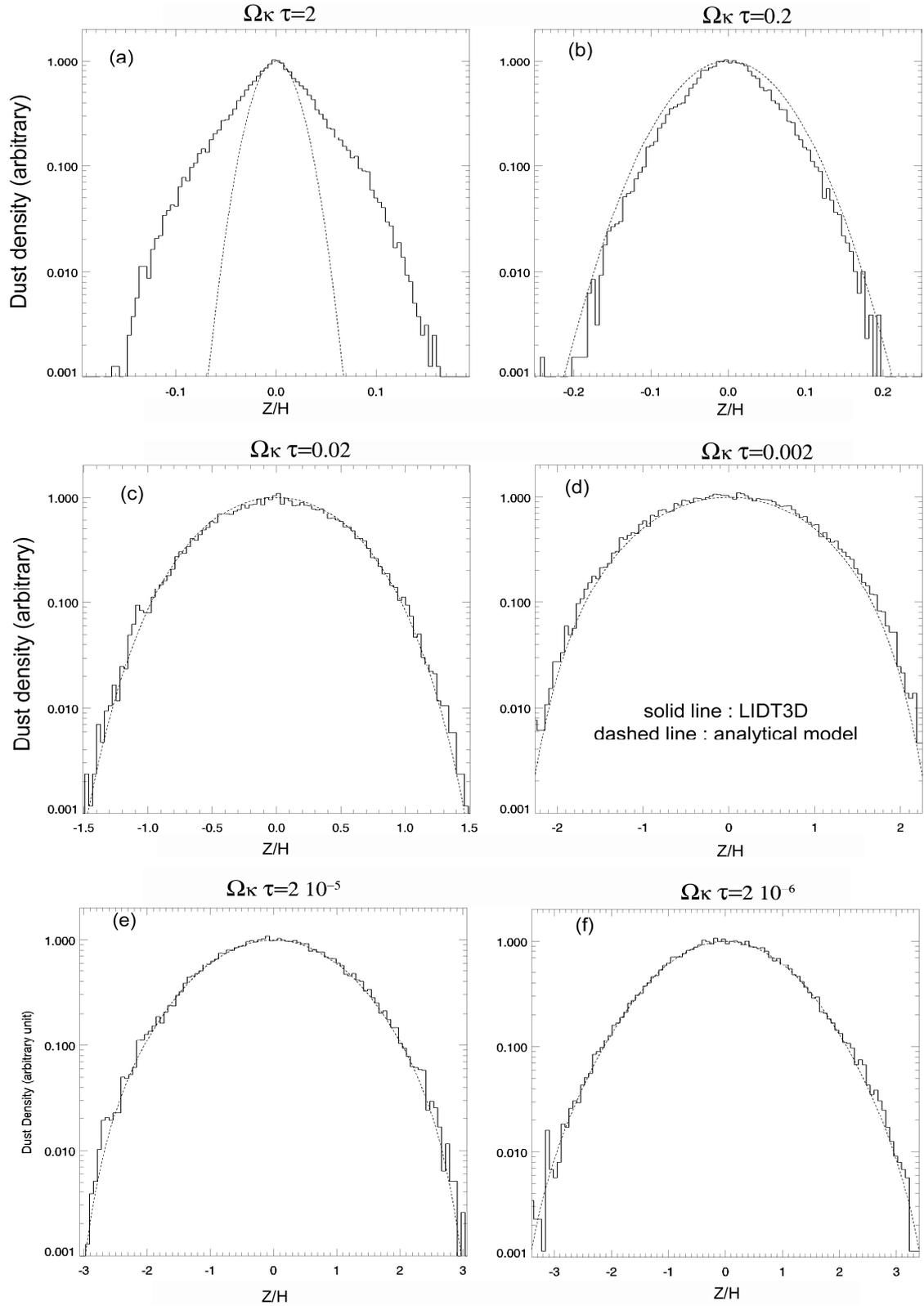

**Figure 3**: Vertical dust density distribution obtained with a constant diffusion coefficient for different-sized particles after 1,000 years of evolution at 1 AU. Solid line: particle distribution obtained with the LIDT3D code; dashed-line: analytical model for small particles tightly coupled to the gas (Eq. [22]). In each graph the title reports $\tau\Omega$ (i.e the particle stopping time times the local Keplerian frequency), which is roughly the Stokes number. Note the excellent agreement that is obtained for $\tau\Omega_k <0.2$. For larger values of $\tau\Omega_k$ the analytical model for particles tightly coupled to the gas fails. The case of big particles is presented in Figure 4.

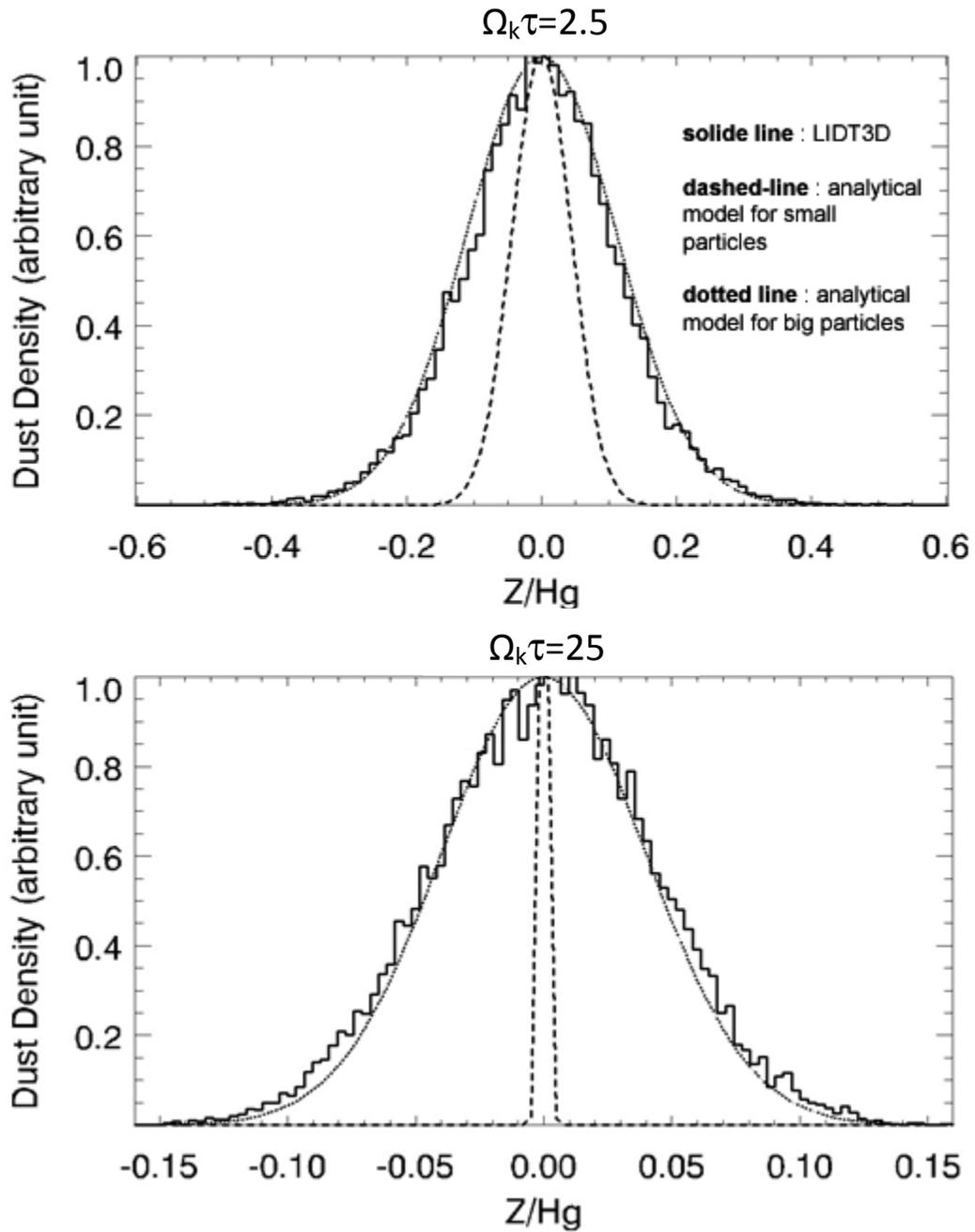

**Figure 4:** Vertical distribution of particles loosely coupled to the gas (τΩ>1, corresponding to big particles). The dashed line shows the expected distribution for particles tightly coupled to the gas, neglecting oscillations, while the dotted line shows the analytical model for particles loosely coupled to the gas (see Eq. [23]), which shows an excellent agreement with LIDT3D. Note that the particle's scale height diminishes with the particle's stopping time in agreement with Youdin and Lithwick (2007).

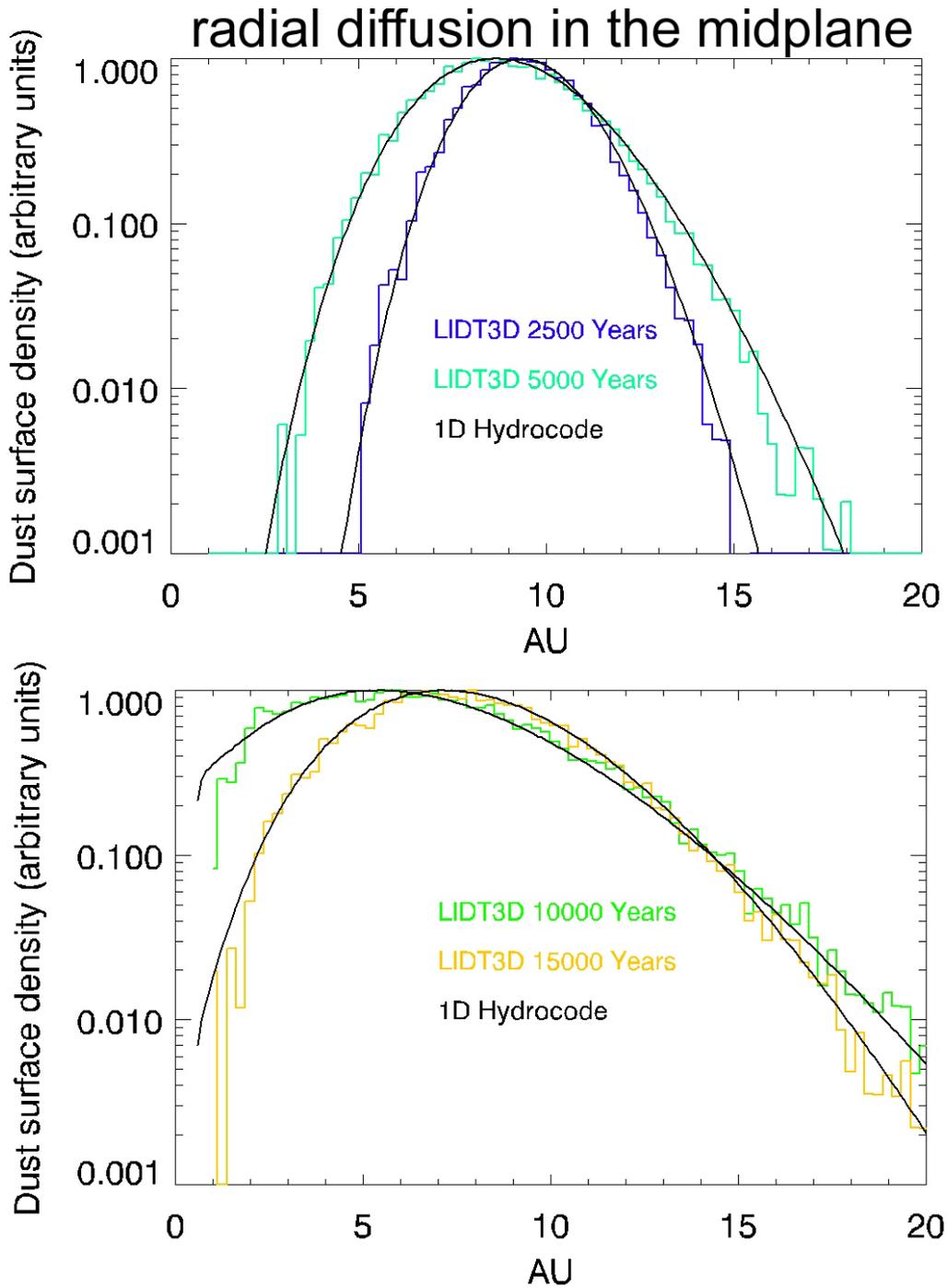

**Figure 5**: Test of radial diffusion. 10,000 particles were released at 10 AU initially, with a 0.1-micron radius. They are forced to evolve in the midplane only of a gaseous disk with a gas surface density decreasing as $r^{-1}$. These plots show the dust surface density after 2,500 and 5,000 years (top) and 10,000 and 15,000 years (bottom) evolution. For comparison, the surface density obtained from the explicit integration of the 1D hydrodynamical diffusion Eq. (23) is shown in black and shows excellent agreement with the results of LIDT3D.

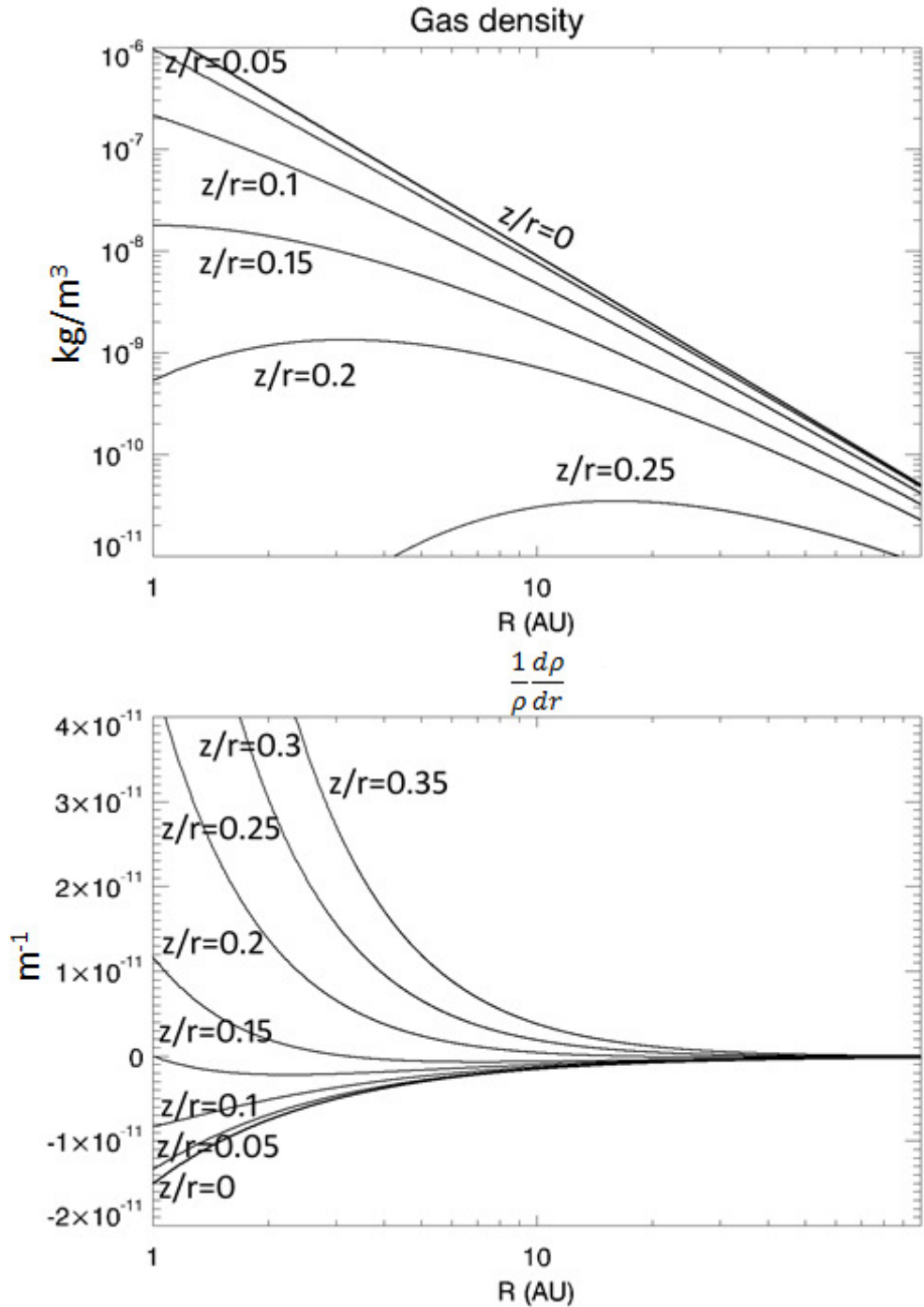

**Figure 6**: Top: Gas density in a disk with a surface density decreasing as $r^{-1}$ for different values of Z/R. Note that the steepest gradient is in the midplane (Z/R=0). Bottom: corresponding values of $1/\rho_g \times \text{grad}(\rho_g)$, which gives the net direction of the local diffusive flux. Note that in the midplane (Z/R=0) the diffusive flux is always negative (i.e. directed inwards), whereas above the midplane, for Z/R>0.15 the diffusive flux can be positive (i.e., directed outwards).

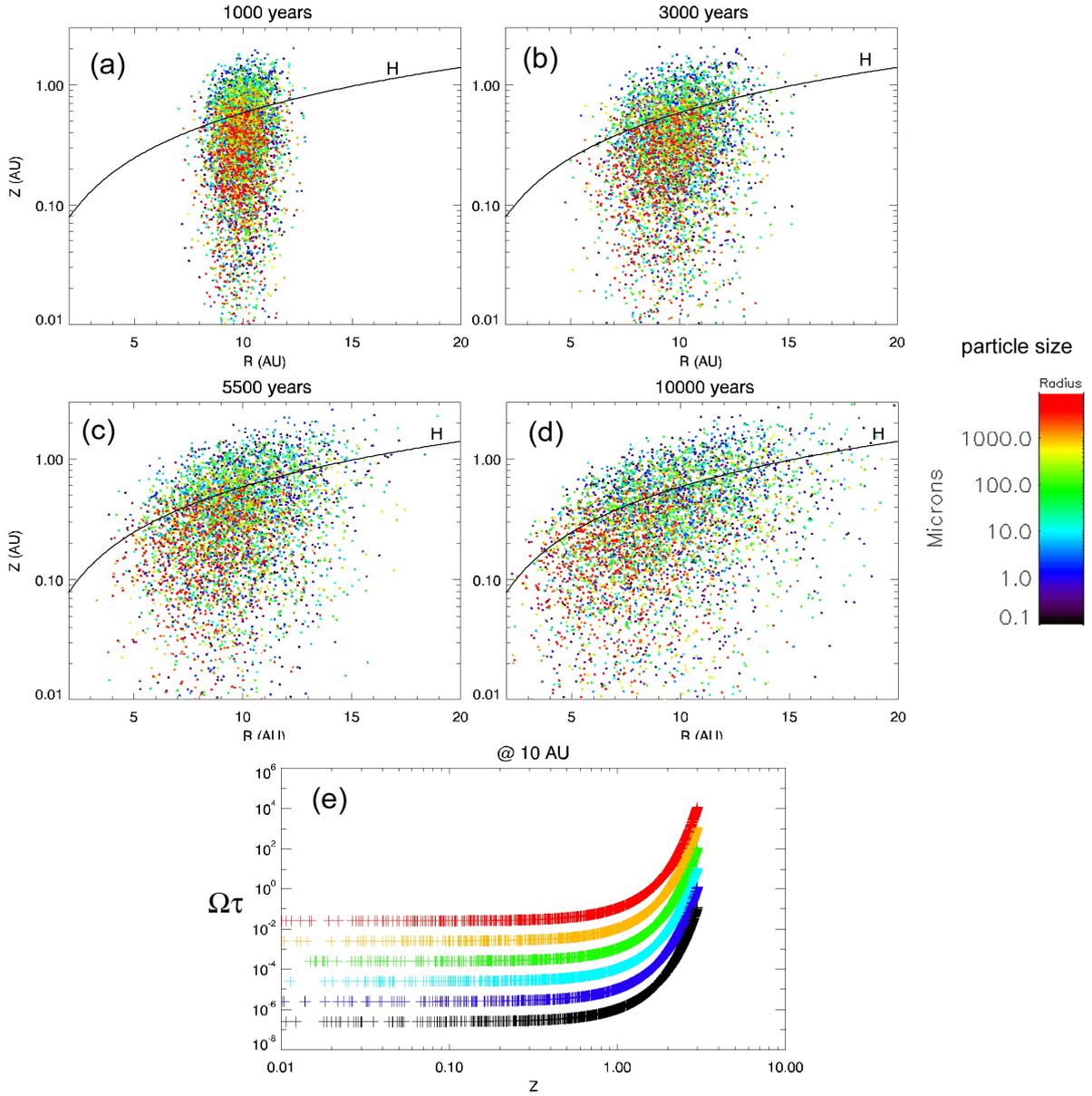

**Figure 7:** Example of diffusion at 10 AU in the (R,Z) plane. 5,000 particles of different sizes (see color scale on the right) were released at 10 AU. For 500 years they do not evolve radially in order to reach a vertical steady state. Then they are allowed to evolve freely in the turbulent gas-disk. In Figures (a) to (d) the (R,Z) location of all particles is displayed. The solid line shows the pressure scale height. Here the Schmidt number is taken from Youdin and Litchwick (2007). Note that radial diffusion is more efficient at higher values of Z/H. In Figure (e) the particle-gas coupling parameter, $\Omega\tau$ (~Stokes number), is displayed as a function of Z at 10 AU, and for different particles sizes (colors).

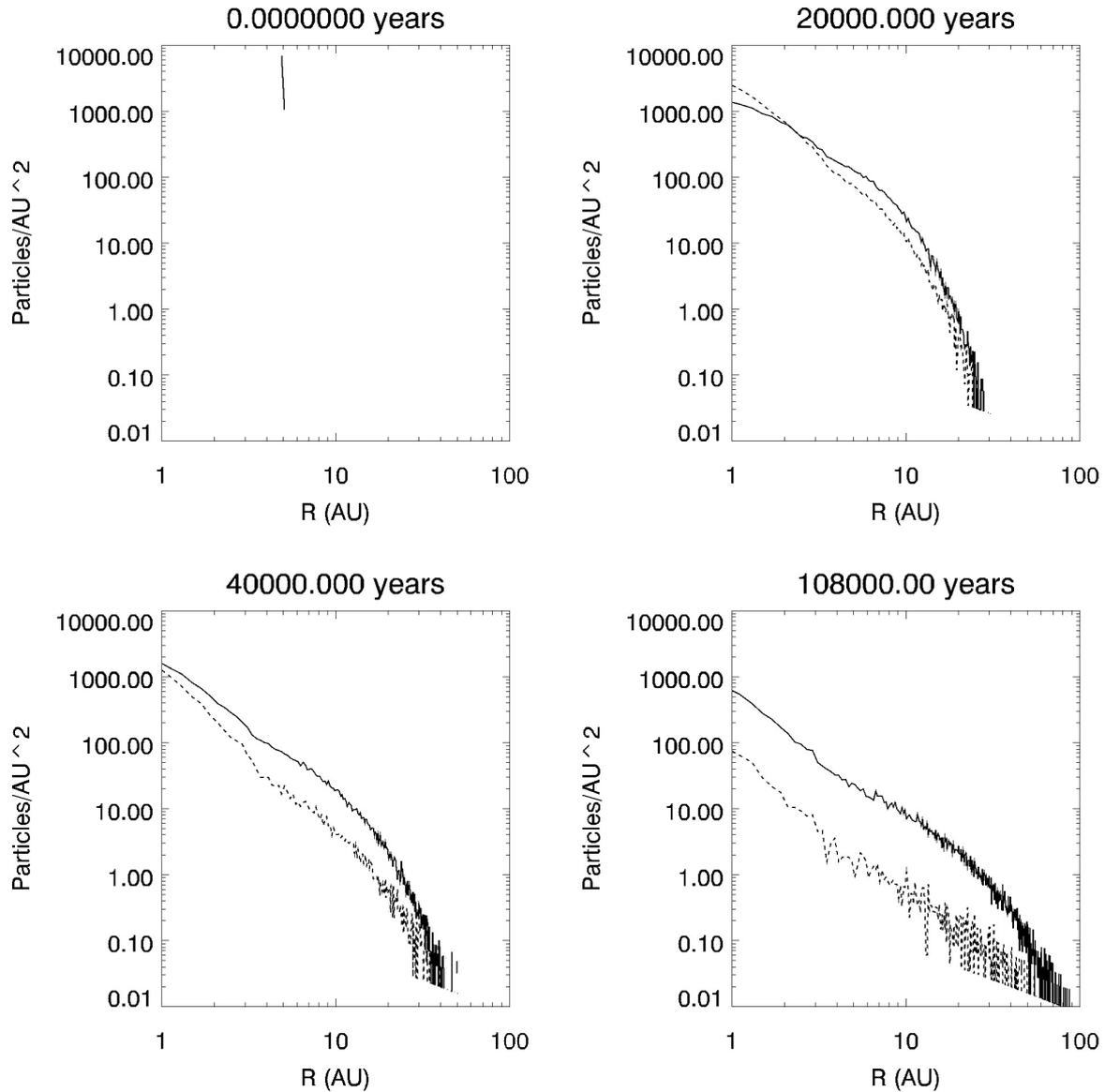

**Figure 8:** Comparison of midplane and 3D diffusion. This figure displays the surface density of particles as a function of the distance to the star at four different epochs. The disk has a surface density decreasing as $r^{-1}$ and $\alpha=0.01$. Solid line: 3D simulation; dashed line: 2D simulation in which particles are confined to the disk's midplane. The particle's radius is 0.1 micron.

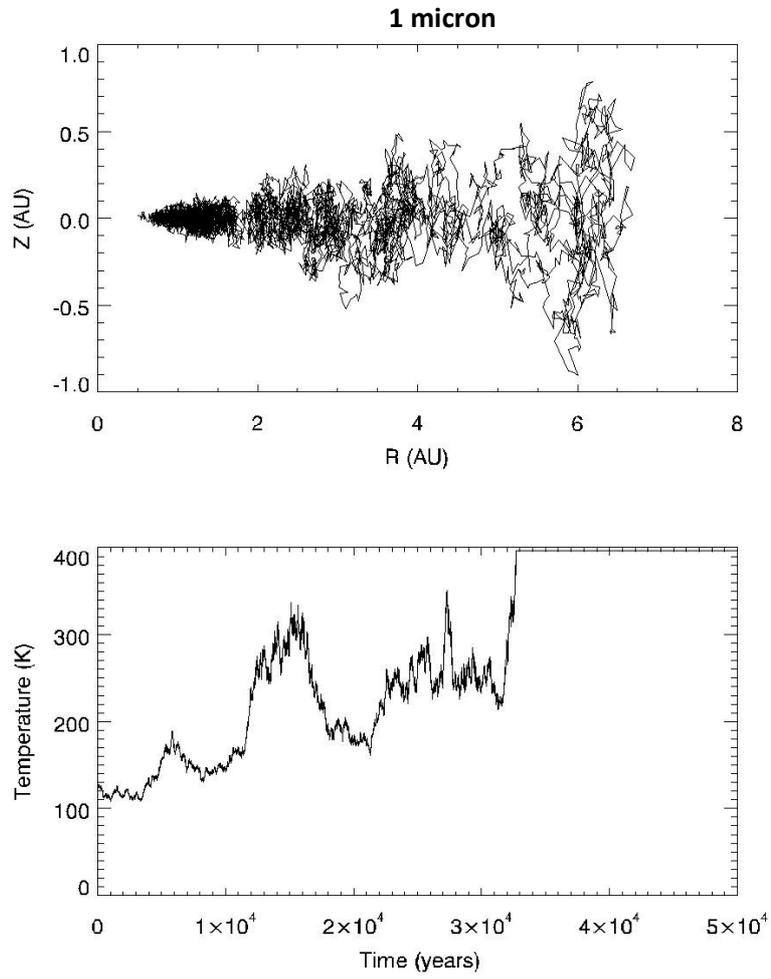

**Figure 9**: Example of the trajectory of a 1-micron particle. Top: path in the R/Z plane; bottom: temperature of the gas surrounding the particle as a function of time. When the particle falls below 0.5 AU (~33,000 years here), it is eliminated from the disk.

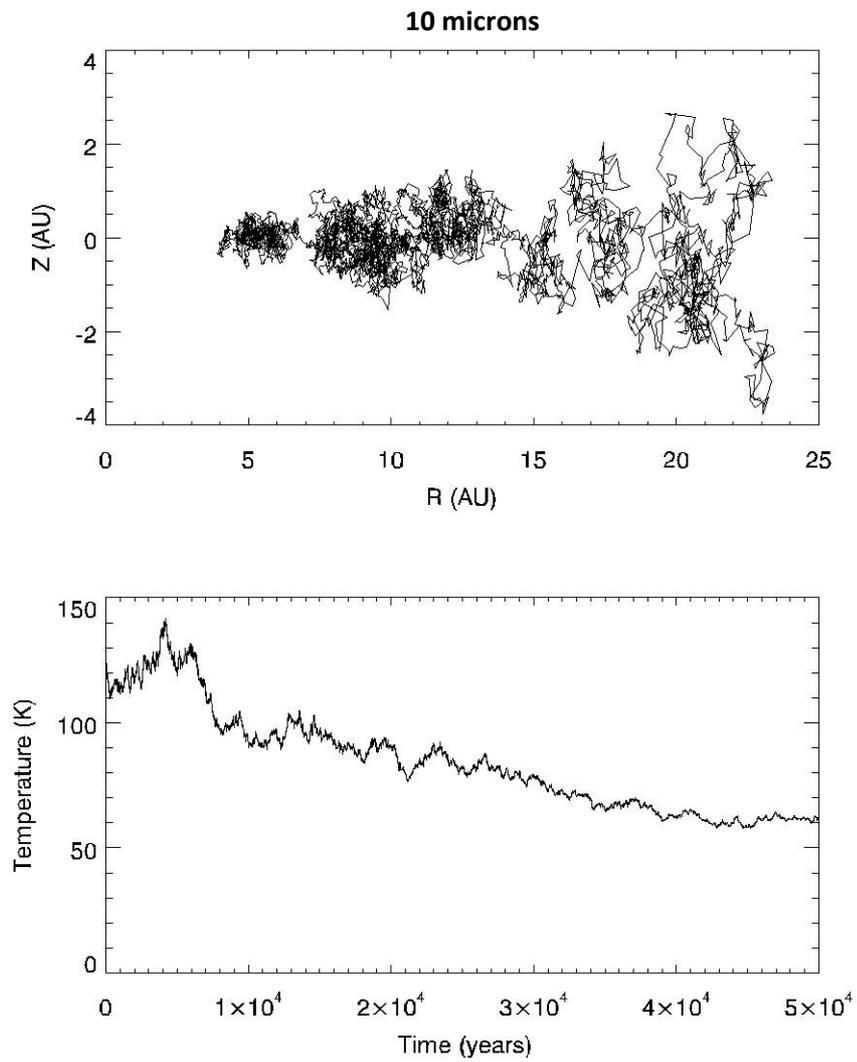

**Figure 10**: Example of the trajectory of a 10 micron particle. Top: path in the R/Z plane; bottom: temperature of the gas surrounding the particle as a function of time.

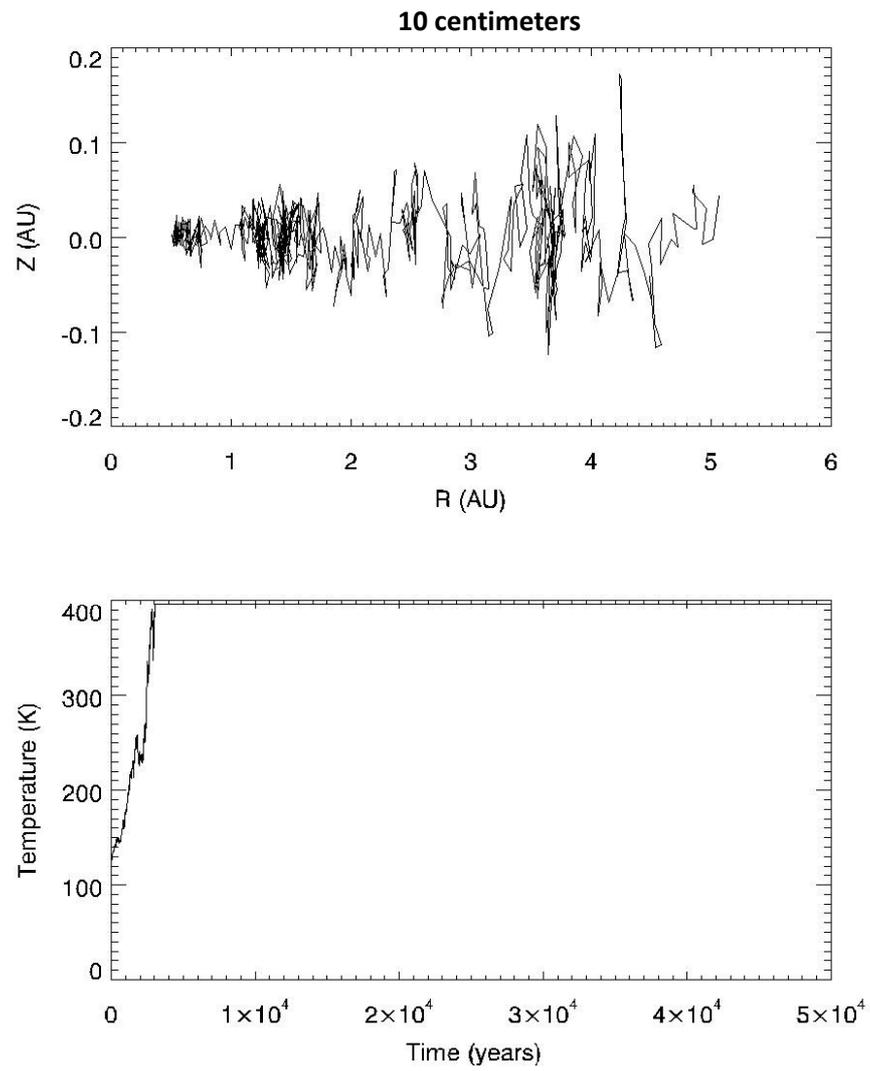

**Figure 11**: Example of the trajectory of a 10-centimeter grain. Top: path in the R/Z plane; bottom: temperature of the gas surrounding the particle as a function of time.

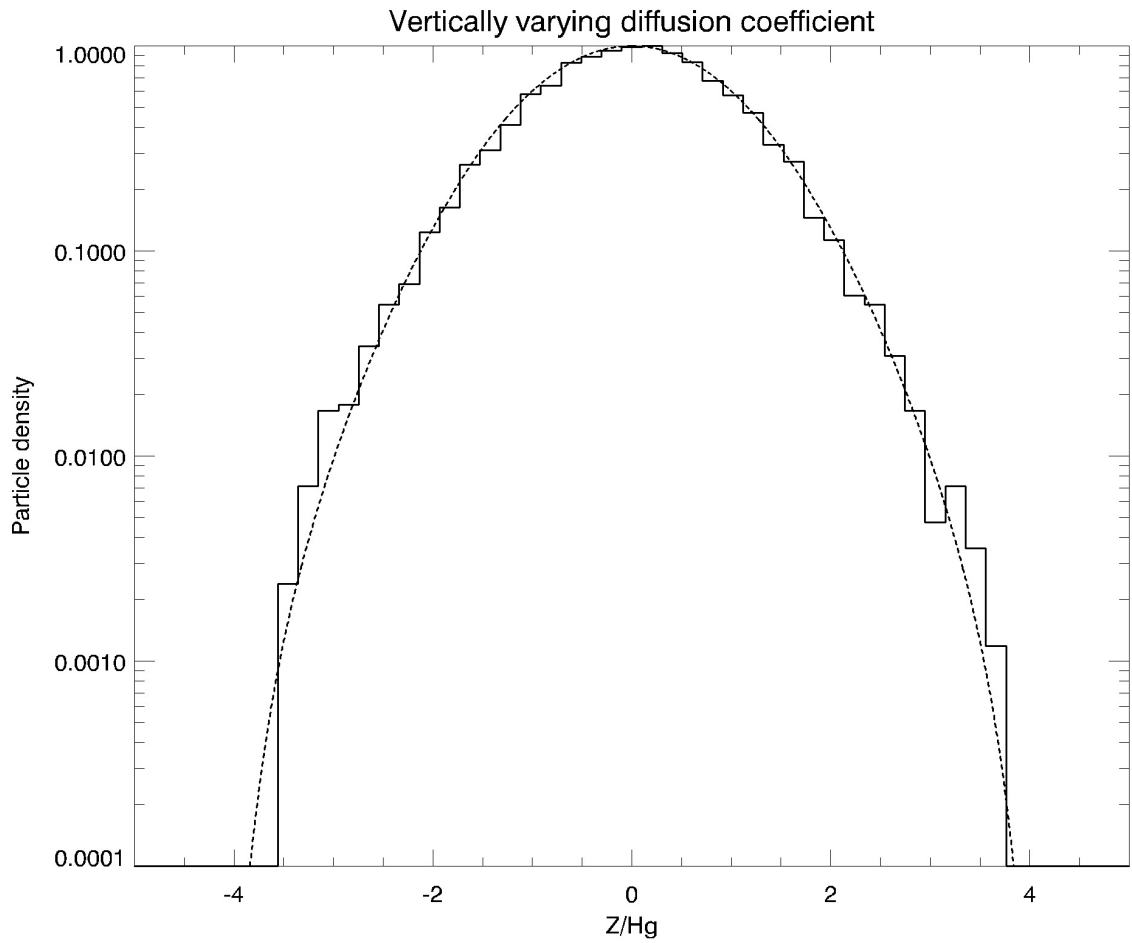

**Figure A1:** Distribution of particles at steady-state in the case of a diffusion coefficient that depends on Z (Eq. 29). We performed a run with $10^4$ particles with 0.1 micron radius particles at 1 AU. Dashed line: analytical steady state solution (Eq. A12 and A13), solid line: particle density obtained in the simulation. See Annex 2 for more details.